\documentclass[preprint,12pt,authoryear]{elsarticle}

\usepackage{amssymb}
\usepackage{amsmath}

\usepackage{bm}
\usepackage{longtable}

\usepackage{color}

\journal{Icarus}

\begin{document}

\begin{frontmatter}


 \author{Kenji Kurosaki\corref{cor1}\fnref{label0}}
 \ead{kkurosaki@nda.ac.jp}
 \cortext[cor1]{Corresponding author}

\title{Reaccumulation process after a catastrophic disruption event on a differentiated asteroid} 


\author[label1]{Masahiko Arakawa} 

\affiliation[label0]{organization={Department of Applied Physics, National Defense Academy},
            addressline={1-10-20, Hashirimizu}, 
            city={Yokosuka},
            postcode={239-8686}, 
            state={Kanagawa},
            country={Japan}}

\affiliation[label1]{organization={Graduate School of Science, Kobe University},
            addressline={1-1, Rokkodaicho, Nada}, 
            city={Kobe},
            postcode={657-0017}, 
            state={Hyogo},
            country={Japan}}

\begin{abstract}
Rubble-pile asteroids can form through the self-gravitational reaccumulation of fragments produced during large-scale collisions.
To investigate how differentiated bodies are disrupted and how iron-rich rubble piles may form, we performed smoothed particle hydrodynamics simulations of impacts between differentiated asteroids with molten or solidified interiors.
Our results show that catastrophic disruption produces a sheet-like structure in which core and mantle materials are stretched and subsequently fragment under self-gravity.
The resulting fragments exhibit nearly identical iron-rock mass ratios, indicating that catastrophic disruption naturally generates numerous compositionally similar fragments.
The largest remnant formed in such events is therefore an iron-rich rubble pile assembled from these mixed fragments, whereas remnants formed through mantle stripping retain a layered structure with an iron core and rocky mantle.
We further find that fragment production is sensitive to material strength and the equation of state: mantle strength reduces the number of small fragments, while core strength suppresses catastrophic disruption when the core is solid.
These results imply that iron-rich rubble-pile asteroids can form only when the iron core is molten.
Our findings provide a unified framework for the formation of metal-rich asteroids such as (16) Psyche and the (22) Kalliope system, and offer predictions for the surface and internal structure that the NASA Psyche mission may test.
\end{abstract}




\end{frontmatter}


\section{Introduction}
The spectral types of asteroids are thought to be closely related to the compositions of meteorites. Asteroid taxonomy has been widely based on the Tholen system \citep{Tholen1984} and the Bus-DeMeo system \citep{Bus2002,DeMeo2009}.
Among the various spectral classes, M/X taxonomic class are of particular interest.
Polarimetric data suggest that their surfaces are covered with Fe-Ni-rich powdered material, indicating that M-type asteroid may be the parent bodies of iron meteorites \citep{Dollfus1979,Hardersen2005,Cloutis2010,Mahlke2022,Wei2025}.

Iron meteorites are key samples for understanding the igneous processes that occurred within their parent bodies in the asteroid belt,
and constraining their origins is essential for understanding planetesimal formation and differentiation in the early solar system.
Among these bodies, 16 Psyche is considered one of the largest M-type asteroids based on remote-sensing observations
\citep{deKleer2021,Shepard2021,Cambioni2022,Dibb2023,Dibb2024}.
The NASA Psyche mission is expected to improve our understanding of the surface characteristics of the M-type asteroid, such as the iron-to-rock ratio, detailed spectral type, and boulder sizes \citep[e.g.,][]{Zuber2022,Nichols-Fleming2024}.
The bulk density of 16 Psyche is currently estimated to be approximately 4.2~g/cm$^3$, and its formation processes may imply that iron-rich material reaccumulated after an impact event \citep[e.g.,][]{Farinella1982,Farnocchia2024}.  
Its origin remains uncertain, with competing hypotheses suggesting it may be either a remnant of a differentiated parent body or a metal-rich asteroid that never underwent differentiation \citep[e.g.,][]{Elkins-Tanton2020,Elkins-Tanton2022}.
Furthermore, understanding the properties of metal‑rich asteroids such as (16) Psyche provides valuable insights into the composition and origin of Mercury, which is thought to possess a large metallic core.

Impact events among asteroids are essential physical processes that create meteorites composed of impact-induced fragments.
Iron meteorites are understood to be fragments formed from differentiated asteroids \citep{Burbine2002,Kleine2005,Bottke2006,Elkins-Tanton2011,Jutzi2013}.
Knowledge of the origin of iron meteorites is essential to investigate the impact and reaccumulation of differentiated asteroids.
Iron-rich planets, especially Mercury, could formed by impact stripping of the planetary mantle \citep[e.g.,][]{Benz1988,Benz2007,Marcus2009,Asphaug2014,Chau2018,Carter2018,Franco2022,Dou2024,Cambioni2025}.
On the other hand, the catastrophic disruption of proto-Mercury provides another insight into the formation of iron-rich asteroids.
\citet{Benz1988} calculated the reaccumulation of fragments generated by the catastrophic disruption of proto-Mercury and suggested that iron-rich clumps were formed through the catastrophic disruption, though the size distribution and compositions were not determined due to the lack of resolution.

Previous studies have investigated the fragmentation and reaccumulation processes resulting from either high-velocity impacts \citep[e.g.,][]{Michel2002,Michel2003,Jutzi2019b,Sugiura2020,Jutzi2020}
or tidal disruption events \citep[e.g.,][]{Hyodo2017,Ruiz-Bonilla2021,Kegerreis2025}.
These studies primarily focused on the reaccretion of rocky components and typically considered unmelted small bodies belonging to the C-complex and S-complex asteroid.
Those studies, however, cannot provide insights into the origin of iron-rich asteroids or fragments, because they considered undifferentiated bodies composed of a single material.
Nevertheless, the catastrophic disruption of differentiated bodies comparable in size to (4) Vesta plays an important role in planetary formation processes, including the quantitative assessment of disruptive collisions during planetesimal accretion and the effects of mantle stripping and fragment reaccumulation on the compositional evolution of solid materials.
\citep{Jutzi2015,Emsenhuber2018,Emsenhuber2024,Sugiura2022,Gabriel2023,Shuai2024, Walte2023,Kobayashi2026}.
Both \citet{Cambioni2026} and \citet{Shuai2024} investigated not only the largest remnant but also the mass and compositional evolution of the surviving projectile.
In addition, \citet{Cambioni2026} also considered collisions during the molten phase of planetesimals.
However, these studies did not examine the possibility of forming metal-rich rubble-pile asteroids, because molten fragments re-equilibrate into a core-mantle structure and solid-core impacts were not modeled with material strength. Our work therefore complements these studies by demonstrating that, once the parent body has solidified, catastrophic disruption can produce metal-rich rubble-pile asteroids, particularly in low-angle collisions.
Although such smaller fragments, particularly those associated with Psyche may not correspond directly to currently observed asteroid families \citep[e.g.,][]{Nesvorny2015}, they could still serve as the parent bodies of iron meteorites, making their characterization potentially important.

Our study examines the catastrophic disruption of a differentiated asteroid, consisting of both mantle and core materials.
We show that during such an event, mantle rock tends to escape, while denser core material is more likely to reaccumulate.
This multi-component reaccretion process provides a new perspective on the origin of iron-rich fragments and may offer insights into the formation of iron-rich asteroids such as 16 Psyche.
In this study, we focus particularly on collisions between fully molten bodies occurring within the first $\sim 100$ Myr after their formation after melting due to radiogenic heating \citep[e.g.,][]{Dodds2021}, and we assume a scenario in which such molten bodies undergo catastrophic disruption followed by reaccretion. Because these bodies are fully molten, we assume that they possess no material strength.
For comparison, we also investigate cases in which either the rocky mantle or the metallic core has solidified, allowing us to assess how material strength influences catastrophic disruption and the resulting fragment population. Here, we analyze both the mantle-stripping process and the reaccumulation of impact-generated debris following the disruption event.

The paper is organized as follows. 
The numerical model and settings are described in Section~\ref{method}.
We present the impact simulation and reaccumulation results in Section~\ref{result}.
We discuss compositions of reaccumulated fragments in Section~\ref{discussion}.
Finally, we summarize the findings in Section~\ref{conclusion}.

\section{Methods} \label{method}
In this study, we perform two types of simulations: a fully molten, strengthless model (\S\ref{hydr}), and strength-included models in which the mantle is solid while the core is either molten or solid (\S\ref{matmodel}).
In section~\ref{sub_example}, \ref{mantle_impct}, and \ref{catastro_imp} results are hydrodynamic simulation described in section~\ref{hydr}.
In section~\ref{dis:mat-chk} and \ref{dis:valid} results are considered the material strength described in section~\ref{matmodel}.

\subsection{Hydrodynamic simulation}\label{hydr}
We performed numerical simulations of the impact event on differentiated asteroids using the standard smoothed particle hydrodynamics method, hereafter SPH \citep[e.g.,][]{Monaghan1992}. 
The standard SPH method is widely used for astronomical collisions.
We use a three-dimensional smoothed particle hydrodynamics impact code developed in \cite{Kurosaki2019,Kurosaki2023}. 

In our simulation, we use the smoothed hydrodynamics simulation \citep[e.g.,][]{Lucy1977,Monaghan1992} 
to solve the following hydrodynamic equations:
\begin{eqnarray}
\frac{d\rho}{dt} &=& -\rho \nabla\cdot\bm{v} \\
\frac{d\bm{v}}{dt} &=& -\frac{1}{\rho} \nabla P + \nabla \int dx'^3 \frac{G\rho(x')}{|\bm{x}-\bm{x'}|} \\
\frac{du}{dt} &=& -\frac{P}{\rho} \nabla \cdot\bm{v} \\
P &=& P(\rho, u)  \label{eos}
\end{eqnarray}
where $\rho, P, \bm{v}$ and $u$ are density, pressure, velocity, and specific internal energy, respectively.
$t$ is the time, $\bm{x}$ is the position, and
$G(=6.67408\times10^{-8}~\mathrm{cm}^3~\mathrm{g}^{-1}~\mathrm{s}^{-2})$ is the gravitational constant.
We normalize the overall impact event by using the target's free fall time described as
\begin{equation}
\tau_\mathrm{ff} = \sqrt{\frac{3\pi}{32 G \rho_T}}, \label{tauff-def}
\end{equation}
where $\rho_T$ is the mean density of the target. We calculate $t=50~\tau_\mathrm{ff}$.
We have implemented the acceleration modules for our SPH code with FDPS \citep{Iwasawa2016} and FDPS fortran interface \citep{Namekata2018}. 
We set the mass of one SPH particle $10^{-5}~M_\mathrm{target}$.
And the rock particle mass is the same as that of the iron particle mass.
For an example, for $10^{23}$ g target with 30~\% of the iron core we use $3 \times 10^{4}$ particles for $0.3\times10^{23}$ g iron core and $7 \times 10^{4}$ particles for $0.7\times10^{23}$ g rock mantle.
In our simulations, the mass per SPH particle is approximately $10^{18}$ g when using $10^{5}$ particles, and the corresponding smoothing length is about $4\times10^{5}$ cm.
This implies that each particle represents a region larger than 1 km in size.

The rock and iron equations of states are used for the Tillotson equation of state for the basalt and the iron, respectively \citep{Tillotson1962,Melosh1989,Benz1999}. 
We set a target of 10$^{23}$ g that has an iron core surrounded by a rock mantle. 
The mass of the impactor composed of a molten basalt is set to be $10^{21}$ or $10^{22}$ g, 
while the differentiated impactor composed of an iron core surrounded by a basalt molten mantle is $10^{23}$ g.
This study ignores the cohesion and shear strength.
The mass ratios of the iron cores were assumed to be 0\%, 30\%, 50\%, 70\%, and 100\%.
The impact velocities ($v_\mathrm{imp}$) were 0.2 -- 5 km s$^{-1}$.
The impact angles $\theta$ were 0$^\circ$ (head-on) and 45$^\circ$ (oblique). 
The target mass of $10^{23}$ g was chosen to represent a differentiated asteroid comparable in size to Vesta $(\sim 2\times 10^{23}~\mathrm{g})$, which serves as a realistic example of a large, well-studied body in the asteroid belt. 
Our aim was to investigate the catastrophic disruption of such a body and examine the potential for iron-rich reaccumulation.
By focusing on a Vesta-scale target, we ensure that the disruption process remains in the gravity-dominated regime, where our model assumptions are valid.
We then analyze the resulting fragment distribution, particularly in the $10^{21} \text{--} 10^{22}$ g range, which corresponds to the estimated masses of bodies such as (21) Lutetia \citep[][]{Sierks2011}, (16) Psyche \citep{Farnocchia2024}, and (22) Kalliope \citep[][]{Vernazza2021}.
This allows us to explore whether such iron-rich fragments could plausibly originate from the disruption of a larger differentiated parent body.

The specific impact energy $Q_R$ is described in
\begin{equation}
    Q_R = \frac{M_T m_\mathrm{imp}v_\mathrm{imp}^2}{2M_\mathrm{tot}^2} f_M(b), \label{QR}
\end{equation}
where $M_T$ is the target mass, $m_\mathrm{imp}$ is the impactor mass, and $M_\mathrm{tot}$ is the total mass described by $M_\mathrm{tot} = M_T + m_\mathrm{imp}$.
$f_M(b)$ is the geometric overlap factor for the impact parameter $b$ introduced by \cite{Kegerreis2020a,Kegerreis2020b} based on the fractional volume of the target and impactor that interact in a grazing collision, where
$f_M(b)=\frac{1}{4}\frac{(R_T+R_I)^3}{R_T^3+R_I^3}(1-b)^2 (1+2b)$ and
$b=\sin\theta$ for the impact angle $\theta$ and the impactor radius $R_I$.
We investigate the catastrophic disruption energy of the whole body and the rock mantle. 
We defined the velocity of escaping particles as faster than the escape velocity calculated by the gravitational potential from the largest fragment \citep[e.g.,][]{Benz1988,Kurosaki2023}.

\subsection{Effect of the material strength}\label{matmodel}
In our simulations, material strength is included by considering two configurations: a solid rock mantle with a molten iron core, and a solid rock mantle with a solid iron core.
Our simulation code for SPH considering the material strength is described \citet{Sugiura2018} and \citet{Kurosaki2026}.
The equations for an elastic body are as follows:
\begin{eqnarray}
    \frac{d\rho}{dt} &=& -\rho\frac{\partial v^\alpha}{\partial x^\alpha} \label{EOC} \\
    \frac{d\bm{v}^\alpha}{dt} &=& \frac{1}{\rho}\frac{\partial \sigma^{\alpha\beta}}{\partial x^\beta} + \nabla \int dx'^3 \frac{G\rho(x')}{|\bm{x}-\bm{x'}|} \label{EOM} \\
    \frac{du}{dt} &=& \frac{1}{\rho}\sigma^{\alpha\beta}\frac{\partial v^\alpha}{\partial x^\beta} \label{EOE}
\end{eqnarray}
where the stress tensor is represented as:
\begin{equation}
    \sigma^{\alpha\beta} = -P \delta^{\alpha\beta} + S^{\alpha\beta}. \label{stress}
\end{equation}
$t$ is the time, $\rho$ is the density, $\bm{v}$ is the velocity vector, $\bm{x}$ is the position vector, $\sigma^{\alpha\beta}$ is the stress tensor, $u$ is the specific internal energy, $p$ is the pressure, $S^{\alpha\beta}$ is the deviatoric stress tensor, $g^\alpha$ is the gravity for $\alpha$ component, and $\delta^{\alpha\beta}$ is the Kronecker delta.
Superscripts indicate directions or components of a vector or tensor, where $\alpha, \beta, \gamma=x,y,z$. 
The summation rule over repeated indices in superscripts, indicated by Greek letters, is applied.
For details on extending SPH to elastic materials.

In our SPH formulation, the deviatoric stress tensor of the $i$-th particle $S_i^{\alpha\beta}$ is calculated by Hook's law represented by
\begin{equation}
    \frac{dS_{i}^{\alpha\beta}}{dt} = 2\mu \left( \dot{\varepsilon}_{i}^{\alpha\beta} - \frac{1}{3}\dot{\varepsilon}_{k}^{\gamma\gamma}\delta_{i}^{\alpha\beta} \right) + S_i^{\alpha\gamma}R_i^{\beta\gamma} + S_i^{\beta\gamma}R_i^{\alpha\gamma}
\end{equation}
where $\mu_i$ is the shear modulus of the $i$-th particle, $\dot{\varepsilon}_{i}^{\alpha\beta}$ and $R_i^{\alpha\beta}$ are a strain rate tensor and a rotational rate tensor of the $i$-th particle, respectively, and are represented as
\begin{eqnarray}
    \dot{\varepsilon}_i^{\alpha\beta} = \frac{1}{2}\left( \frac{\partial v_i^\alpha}{\partial x_i^\beta} + \frac{\partial v_i^\beta}{\partial x_i^\alpha} \right), \\
    R_i^{\alpha\beta} = \frac{1}{2}\left( \frac{\partial v_i^\alpha}{\partial x_i^\beta} - \frac{\partial v_i^\beta}{\partial x_i^\alpha} \right).
\end{eqnarray}
The rock strength model is used for \citet{Collins2004}.
The yield strength model of rock is
\begin{equation}
    Y=Y_\mathrm{d} D + (1-D) Y_\mathrm{i} \label{RYrock}
\end{equation}
where $Y_\mathrm{d}$ is the damaged material strength, $Y_\mathrm{i}$  is the material strength for intact rock, and $D$ is the scalar measure of damage that is computed by the damage model implemented by \citet{Benz1999}
The yield strength of intact rock is defined as
\begin{equation}
    Y_\mathrm{i} = Y_0 + \frac{\mu_\mathrm{i} P}{1+\mu_\mathrm{i} P/(Y_M - Y_0)} \label{RYint}
\end{equation}
where $Y_0$ is the cohesion of the intact rock, $\mu_i$ is the coefficient of internal friction, $Y_M$ is the von Mises plastic limit, respectively.
The damaged rock strength is
\begin{equation}
    Y_\mathrm{d} = Y_\mathrm{d0}+\mu_\mathrm{d} P \label{RYdam}
\end{equation}
where $Y_\mathrm{d0}$ is the cohesion of the damaged rock and $\mu_\mathrm{d}$ is the coefficient of the internal friction for damaged rock, respectively.
Note that $Y_d$ is limited to $Y_\mathrm{d}\le Y_\mathrm{i}$.
We use the Grady-Kipp fragmentation model for rock mantle \citep{Grady1980}, which simulates the activation and growth of flaws in a brittle material \citep{Benz1994,Benz1995}.
Thermal softening is used for \citet{Ohnaka1995} as
\begin{equation}
    Y \to Y\tanh\left( \zeta \left( \frac{T_M}{T}-1 \right) \right) \label{RYtherm}
\end{equation}
where $T_M$ is the melt temperature and $\zeta$ is the material constant of the thermal softening which we set to $\zeta=1.2$ \citep[][]{Collins2004,Emsenhuber2018}.
We summarized the material model parameters for rock mantle on Table~\ref{tab:mat-rock}.

The iron strength model is used for Johnson-Cook model \citet{Johnson1983}, which is also used for the impact cratering simulation on Asteroid (16) Psyche \citep[e.g.,][]{Raducan2020}.
The Johnson-Cook strength model is defined as
\begin{equation}
    Y=(A+B\varepsilon^N) (1+C\ln \dot{\varepsilon})\left[ 1-\left( \frac{T-T_\mathrm{ref}}{T_m - T_\mathrm{ref}} \right)^M \right] \label{JCmodel}
\end{equation}
where $A$ is the yield strength at the reference state, $B$ is the Johnson-Cook coefficient, and $C$ is the strain rate constant, $T$ is the temperature, $T_m$ is the melt temperature, $T_\mathrm{ref}$ is the reference temperature, $n$ is the strain hardening exponent, and $m$ is the thermal softening exponent, respectively.
The model parameter for iron is used \citet{Alexander2022}.
We summarized the material model parameters for iron core on Table~\ref{tab:mat-iron}.
To use the yielding strength, we modify the deviatoric stress tensor $S_i^{\alpha\beta}$ as:
\begin{eqnarray}
    S_i^{\alpha\beta} &\to& f_i S_i^{\alpha\beta}, \\
    f_i &=& \min\left[ \frac{Y}{\sqrt{J_{2,i}}},1 \right],\\
    J_{2,i} &=& \frac{1}{2}S_i^{\alpha\beta}S_i^{\alpha\beta}. \\
\end{eqnarray}

\begin{table}[htbp]
    \centering
    \begin{tabular}{c|c|c}
        Description & Symbol & Rock mantle  \\
        \hline \hline
        Shear modulus [GPa] & $\mu$ & 22.7$^a$  \\
        Cohesion of the intact rock [MPa] & $Y_0$ & 90$^b$ \\
        Coefficient of the internal friction of the intact rock & $\mu_\mathrm{i}$ & 2.0$^b$ \\
        von Mises plastic limit of intact rock [GPa] & $Y_M$ & 1.5$^b$ \\
        Cohesion of the damaged rock [MPa] & $Y_\mathrm{d0}$ & 0.0$^b$ \\
        Coefficient of the internal friction of the damaged rock & $\mu_\mathrm{d}$ & 0.8$^b$ \\
        Melt temperature [K] & $T$ & 1500$^b$ \\
        Thermal softening parameter & $\zeta$ & 1.2$^b$ \\
        Weibull parameter [cm$^{-3}$] & $k$ & $4.0\times 10^{29}$$^a$ \\
        Weibull parameter & $m$ & $9.0$$^a$ \\
        \hline
    \end{tabular} 
    \caption{Material model for basalt rock mantle.$^a$\citet{Benz1999}, $^b$\citet{Collins2004}.}
    \label{tab:mat-rock}
\end{table}

\begin{table}[htbp]
    \centering
    \begin{tabular}{c|c|c}
        Description & Symbol & Iron core  \\
        \hline \hline
        Shear modulus [GPa] & $\mu$ & 82.0 \\
        Yield strength at the reference state [MPa] & $A$ & 397 \\
        Johnson-Cook coefficient [MPa] & $B$ & 861 \\
        Strain-rate constant & $C$ & 0.017 \\
        Strain hardening exponent & $n$ & 0.32 \\
        Thermal softening exponent & $m$ & 0.72 \\
        Melt temperature [K] & $T_\mathrm{ref}$ & 1750 \\
        Reference temperature [K] & $T_\mathrm{ref}$ & 77 \\
        \hline
    \end{tabular} 
    \caption{Material model for Fe-Ni iron core based on \citet{Alexander2022}.}
    \label{tab:mat-iron}
\end{table}

\subsection{Robustness tests} \label{model:av-W}
In SPH simulations, the choice of artificial viscosity and kernel function can also influence the results.
Differences in artificial‑viscosity models may affect the treatment of shock waves \citep[e.g.,][]{Monaghan1983,Hosono2016}, while kernel functions can influence the accuracy of density estimates \citep[e.g.,][]{Swegle1995,Wendland1995,Hongbin2005,Dehnen2012}.
The artificial viscosity based on an analogy with Riemann solvers \citep[e.g.,][]{Monaghan1997,Price2012} has been proposed as a model that more effectively captures shock propagation:
\begin{equation}
  \Pi_{ij}=
  \begin{cases}
    -K\frac{v_{ij}^\textrm{sig}}{\rho_{ij}}w_{ij} & w_{ij} < 0, \\
    0       & \text{otherwise}.
  \end{cases} \label{pairav}
\end{equation}
where
\begin{eqnarray}
    w_{ij} &=& \frac{(\bm{r}_j-\bm{r}_i)\cdot (\bm{v}_j-\bm{v}_i)}{|\bm{r}_j-\bm{r}_i|}, \\
    v_{ij}^\textrm{sig} &=& c_i + c_j -3w_{ij}, \\
    \rho_{ij} &=& \frac{\rho_i + \rho_j}{2}.
\end{eqnarray}
We set $K=0.5$ as a constant value.
For comparison, although we used a Gaussian kernel in the main simulations, we also tested the widely used cubic spline kernel:
\begin{equation}
    W(\bm{r}_{ij},h) = \frac{1}{\pi h^3} 
    \begin{cases}
        1-\frac{3}{2}\left( \frac{|\bm{r}_{ij}|}{h} \right)^2 + \frac{3}{4}\left( \frac{|\bm{r}_{ij}|}{h} \right)^3 & \left( 0\le \frac{|\bm{r}_{ij}|}{h} \le 1 \right) \\
        \frac{1}{4} \left(2- \frac{|\bm{r}_{ij}|}{h} \right)^3 & \left( 1\le \frac{|\bm{r}_{ij}|}{h} \le 2 \right) \\
        0 & \left( 2 < \frac{|\bm{r}_{ij}|}{h} \right)
    \end{cases} \label{ker-S}
\end{equation}

\subsection{Analysis of results}
To compute the cumulative distribution of the number and iron-mass fraction,
we bin the fragments into 40 logarithmically spaced mass bins over the range $M/M_{\mathrm{tot}}=10^{-5}$--1.
The iron-mass fraction in each bin is calculated as the arithmetic mean of the fragments contained in that bin, and the error bars represent the 1--$\sigma$ standard deviation.
To prevent visual overlap of error bars, a small horizontal jitter is applied while ensuring that each point remains within its corresponding mass-bin range.
The cumulative number distributions are computed using the same mass bins, allowing direct comparison between the two panels.

\section{Results} \label{result}
A summary of the results of our numerical simulations is shown in Tables~\ref{tab:result} and \ref{tab:catastro}.
We focus on a target of mass $10^{23}$ g, with the iron core accounting for 30\% and the rock mantle 70\% of the mass.

We show an example of the mantle stripping impact result in Section~\ref{sub_example}.
The mantle critical energy of the mantle stripping is shown in Section~\ref{mantle_impct}.
We also show an example of the catastrophic disruption result in Section~\ref{catastro_imp}.
The cumulative mass distribution and iron mass fraction of fragments generated by the catastrophic disruption are shown in Section~\ref{sub_cum}.

\subsection{Mantle stripping impact} \label{sub_example}
\begin{figure}
    \centering
    \includegraphics[bb=0 0 1920 720,width=\linewidth]{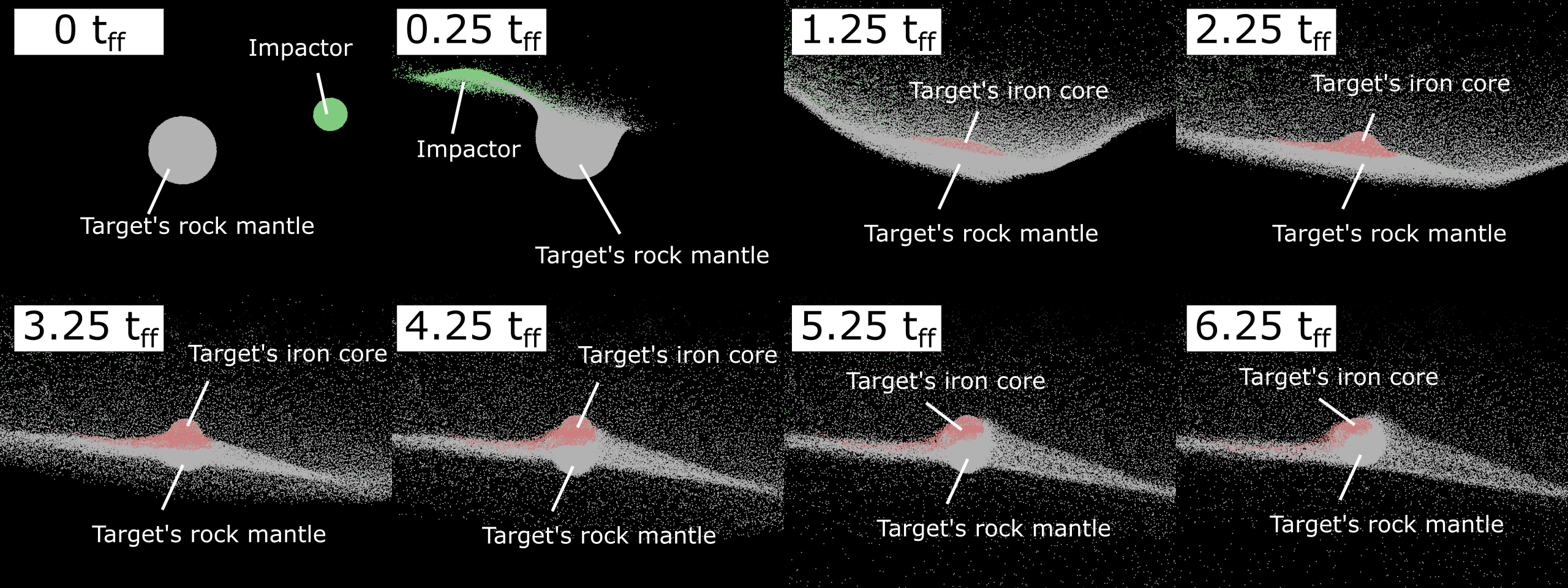}
    \caption{Snapshots of the mantle stripping impact. This simulation shows the RUN 54 shown in Table~\ref{tab:result}.
    The target and impactor mass are $10^{23}$ g and $10^{21}$ g, respectively. 
    The target is consisted of 70\% of the basalt rock mantle and 30\% of the iron core.
    The impactor is consisted of 100\% of the basalt rock.
    The material strength of both the target and impactor is ignored.
    White and green dots are the rock material of the target and the impactor, respectively. Red dots are the iron material of the target.
    The elapsed times in the upper panels are 0, $0.25~\tau_\mathrm{ff}$, $1.25~\tau_\mathrm{ff}$, and $2.25~\tau_\mathrm{ff}$ from left to right, respectively.
    Those in the lower panels are $3.25~\tau_\mathrm{ff}$, $4.25~\tau_\mathrm{ff}$, $5.25~\tau_\mathrm{ff}$, and $6.25~\tau_\mathrm{ff}$ from left to right, respectively.}
    \label{fig:impact_example_small}
\end{figure}

Figure \ref{fig:impact_example_small} shows the impact simulation of Run 54, in which an impactor of $10^{22}$ g collides at 5 km s$^{-1}$ with an impact angle of $45^\circ$, where $Q_R = 2.1 \times 10^{9}$ erg g$^{-1}$.
We can see a strong jet structure followed near-total destruction of the impactor and target.
From $t=0.25\tau_\mathrm{ff}$ to $t=3.25\tau_\mathrm{ff}$, the mantle strips from the core which seems to suffer less from the collision.
The mantle stripping shown in this simulation is the key mechanism that separates the mantle from the core.
We also found that the rock ejecta reaccumulated on the largest fragment. The largest fragment retained a layered structure of an iron core surrounded by a rock mantle.

\subsection{Critical impact energy of mantle stripping} \label{mantle_impct}

\begin{figure}
    \centering
    \includegraphics[bb=0 0 960 360, width=\linewidth]{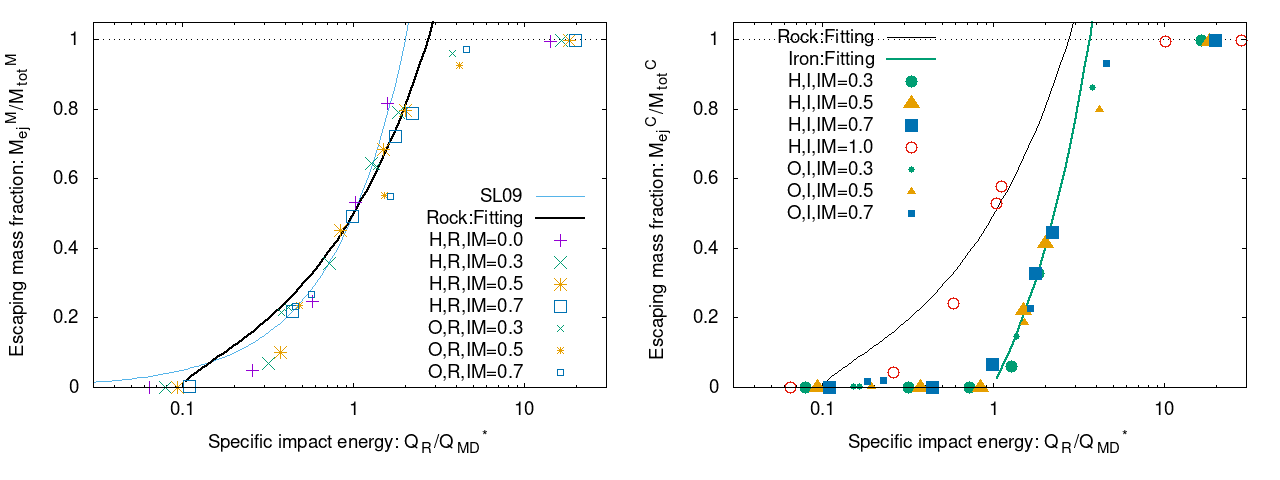}
    \caption{The escaping mass fraction for the rock mantle (left panel) and iron core (right panel) pertaining to the specific impact energy normalized by $Q_\mathrm{MD}^{\ast}$ determined by Table~\ref{tab:qdstar}. "H" denotes the head-on collision, and "O" denotes the oblique collision whose impact angle is $45^\circ$. "R" and "I" denote the escaping mass fraction of the rock $M_\mathrm{ej}^M/M_\mathrm{tot}^{M}$ and the iron $M_\mathrm{ej}^C/M_\mathrm{tot}^{C}$, respectively. "IM" denotes the iron mass fraction of the target. The blue line is a relationship shown by \citet{Stewart2009}. The black and green lines are the fitting relations shown in Eq. \ref{MejR_QD} and Eq.~\ref{MejC_QD}, respectively.}
    \label{fig:IRQMD}
\end{figure}

We next analyze the ejection mass of the mantle stripping shown in Figure~\ref{fig:IRQMD}.
Table~\ref{tab:qdstar} shows the critical impact energy that would remove half the mass of the mantle $Q_\mathrm{MD}^\ast$. 
The critical impact energy that would remove half the total mass $Q_\mathrm{D}^\ast$ is also shown.
The catastrophic disruption energy for the bulk body increases when the target's iron mass fraction increases because the binding energy induced by the gravitational potential of the iron-rich target is greater than that of the iron-poor target due to the difference in their mean densities.
On the other hand, the catastrophic disruption energy for the rock mantle is nearly constant, even when the target's iron mass fraction changes.

$Q_\mathrm{MD}^\ast$ will be helpful in analyzing the mantle stripping.
Figure~\ref{fig:IRQMD} shows the escaping mass fraction of the rock mantle $M_\mathrm{ej}^M/M_\mathrm{tot}^M$ and iron core $M_\mathrm{ej}^C/M_\mathrm{tot}^C$, 
where $M_\mathrm{ej}^M$ is the escaping mass of the rock mantle, $M_\mathrm{tot}^M$ is the total mass of the rock mantle, $M_\mathrm{ej}^C$ is the escaping mass of the iron core, and $M_\mathrm{tot}^C$ is the total mass of the iron core. 
We find that the ejected mass of the rock mantle is expressed by a linear relation of the specific impact energy normalized by the catastrophic disruption energy for the rock mantle $Q_\textrm{MD}^{\ast}$.
\citep[e.g.,][]{Benz1999,Stewart2009,Leinhardt2009,Leinhardt2012,Genda2017}.
$Q_{\mathrm{MD}}^\ast$ is defined as the specific impact energy required for the mass of the target’s rocky mantle to be reduced to half of its initial value.
The iron core did not escape since the rock mantle escaped preferentially when $Q_R < Q_\mathrm{MD}^\ast$.
We derived scaling laws for the escaping mass fractions of the rock mantle $M_{\mathrm{ej}}^{\mathrm{M}}$ and the iron core $M_{\mathrm{ej}}^{\mathrm{C}}$ as functions of the specific impact energy $Q_R$.
Rock mantle ejects only for $Q_R/Q_{MD}^{\ast}>0.1$, whereas iron core ejects for $Q_R/Q_{MD}^{\ast}>1$.
These thresholds are incorporated into the fitting functions:
\begin{eqnarray}
    \frac{M_\mathrm{ej}^\mathrm{M}}{M_\mathrm{tot}^\mathrm{M}} &=& 0.53 \left( \frac{Q_R}{Q_\mathrm{MD}^{\ast}} - 0.1  \right)^{0.65}, \label{MejR_QD} \\
    \frac{M_\mathrm{ej}^\mathrm{C}}{M_\mathrm{tot}^\mathrm{C}} &=& 0.40 \left( \frac{Q_R}{Q_\mathrm{MD}^{\ast}} - 1.0  \right)^{0.94}.\label{MejC_QD}
\end{eqnarray}
The offsets of 0.1 in Eq~\ref{MejR_QD} and 1.0 in Eq.~\ref{MejC_QD} represent the onset of rock mantle and iron core ejection, respectively.
The exponent is slightly smaller than unity, indicating that catastrophic disruption does not scale purely with impact energy but also reflects momentum-dominated fragmentation processes.
These relations are empirical fits to our simulation suite and are not intended as universal scaling laws; rather, they summarize the behavior of differentiated bodies within the parameter range explored in this study.
Eq.~\ref{MejR_QD} describes the stripping of the outer rocky mantle and therefore applies only to material located near the surface.
It does not capture the behavior of the deeply buried iron core, whose ejection depends sensitively on compression wave propagation, and the geometry of the collision.
As a result, the iron core escaping mass fraction does not follow the same scaling relation as the mantle.
Only in the limiting case of a fully iron target (100\% iron) would the core behave analogously to the mantle, but such cases are not not representative of differentiated bodies whose bulk iron content is equal to the bulk value of the solar system \citep[30 wt.\%][]{Palme2003}.
We found that, for $Q_R < 2 Q_\mathrm{MD}^{\ast}$
, the ejected masses of both the rock mantle and the iron core are well described by Eqs. \ref{MejR_QD} and \ref{MejC_QD}, respectively, and are independent of the initial core-mantle fraction and the impactor mass.
On the other hand, the reaccumulation process of fragments generated by the catastrophic disruption occurs when $Q_R \sim Q_\mathrm{MD}^{\ast}$. 
Under that condition, the mass of the largest remnant is less than 10\% of the total mass. 
The largest remnant is formed via the reaccumulation of small fragments shown in Figure \ref{fig:impact_example}.
On the other hand, the reaccumulation of small fragments does not occur efficiently when $Q_R \sim 10 Q_\mathrm{MD}^{\ast}$. 
In the following subsection, we focus on the impact events of $Q_R \sim  Q_\mathrm{MD}^{\ast}$ to discuss the reaccumulation process of fragments.

\subsection{Catastrophic disruption} \label{catastro_imp}
\begin{figure}
    \centering
    \includegraphics[bb=0 0 1920 720,width=\linewidth]{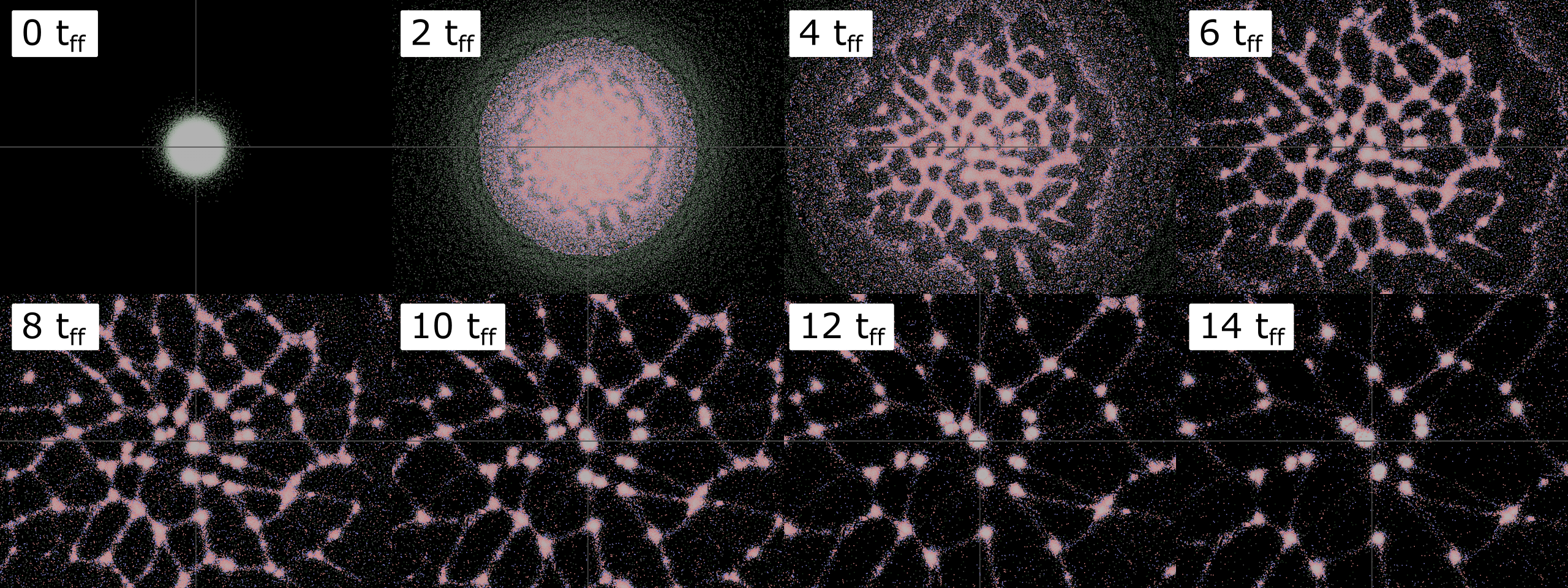}    \caption{Snapshots of the catastrophic disruption viewed from the direction of the collision. This figure shows the result of Run 13 listed in Table \ref{tab:result}. The target has an 70\% of the basalt rock mantle and 30\% of the iron core, and the impactor is identical to the target. The collision is head-on with an impact velocity of 1 km s$^{-1}$. The material strength of both the target and impactor is ignored.
    White and green dots are the rock material of the target and the impactor, respectively. Red dots are the iron material of the target.
    The elapsed times in the upper panels are 0, $2~\tau_\mathrm{ff}$, $4~\tau_\mathrm{ff}$, and $6~\tau_\mathrm{ff}$ from left to right, respectively.
    Those in the lower panels are $8~\tau_\mathrm{ff}$, $10~\tau_\mathrm{ff}$, $12~\tau_\mathrm{ff}$, and $14~\tau_\mathrm{ff}$ from left to right, respectively.}
    \label{fig:impact_example}
\end{figure}

The off-center collision is less effective in ejecting mass from the target than the head-on collision because the jetting mechanisms responsible for escaping mass depend on the compression wave, which is maximized for a head-on collision.
Figure \ref{fig:impact_example} shows the impact simulation of Run 13, which is the head-on collision of the differentiated impactor of $10^{23}$ g colliding at 1 km s$^{-1}$, where $5\times v_\mathrm{esc}$.
When the impactor collides with the target, a strong jetting occurs in a plane almost perpendicular to the collision axis, which is the same phenomenon as shown in \cite{Benz1988}.
The jetting mechanism resembles the condition discussed in a spherical projectile colliding with a flat target \citep[e.g.,][]{Johnson2014,Okamoto2020}.
As the collision proceeds, the cores of the target and impactor are deformed into a thin sheet in the plane perpendicular to the collision axis.
The expansion velocity of the sheet is slow enough to reaccumulate surrounding materials due to their self-gravity.
The inner parts of the sheet make filament structures that produce small clumps.
The clumps forming in the filaments are still gravitationally bound and eventually fall back, coagulating remnants.
Such catastrophic disruption occurs when the impact velocity is faster than 1 km s$^{-1}$, where $Q_R = 1.25 \times 10^9~\mathrm{erg}~\mathrm{g}^{-1}$. 
We find that small iron-rich bodies are formed due to the gravitational reaccumulation process.
Since a significantly greater mass of the rock mantle is ejected compared to that of the iron core, the reaccumulated bodies are expected to be iron-rich. 
Such catastrophic collision results are summarized in Table \ref{tab:catastro}.
\citet{Benz1988} also showed a similar annular structure in their simulations of Mercury-scale impacts, although their results indicated an almost entirely iron core, partly due to numerical‑resolution effects on the spatial distribution of SPH particles.
Figure~\ref{fig:impact_example} shows a head-on collision, which is the most favorable geometry for forming a sheet-like structure through core-core interaction.
Our calculation showed that the impact-generated sheet-like structure was formed when the impact angle was less than $15^\circ$, while at higher angles (e.g., $45^\circ$), such structures are not observed.
Those differences were due to the collision geometry, which determines the efficient core overlaps.
Although low-angle and head-on collisions are less probable, they are the most efficient at disrupting the iron core and producing iron-rich debris.

\subsection{Fragments generated by a catastrophic disruption event} \label{sub_cum}
In this section, we analyze the short-term reaccretion immediately after the catastrophic disruption. Our SPH simulations follow the evolution up to $t=50 \tau_\mathrm{ff}$.
This allows us to quantify the mass and composition of fragments produced directly by the impact.
The long-term orbital evolution of fragments, typically modeled by coupling SPH with subsequent N-body simulations \citep[e.g.,][]{Durda2007,Benavidez2012,Jutzi2019b}, is beyond the scope of this study. 
Processes such as fragment-fragment collisions and late-time reaccretion are therefore not captured, and the final mass of the largest remnant may be underestimated.
Here we focus on the initial population of iron-rich fragments generated immediately after the disruption, while noting that their long-term dynamical evolution requires future N-body calculations.

Fragments are identified using a clump-finding algorithm with a linking length equal to the smoothing length.
A clump is defined as a group of at least 10 particles.
Clumps whose relative velocity exceeds the escape velocity of the largest remnant are classified as escaping fragments.
The resolution dependence of the clump-finding results is discussed in Section \ref{dis:cd}; while small fragments are resolution-sensitive, the largest fragments are robust across different resolutions.

\begin{figure}
    \centering
    \includegraphics[bb=0 0 900 375,width=\linewidth]{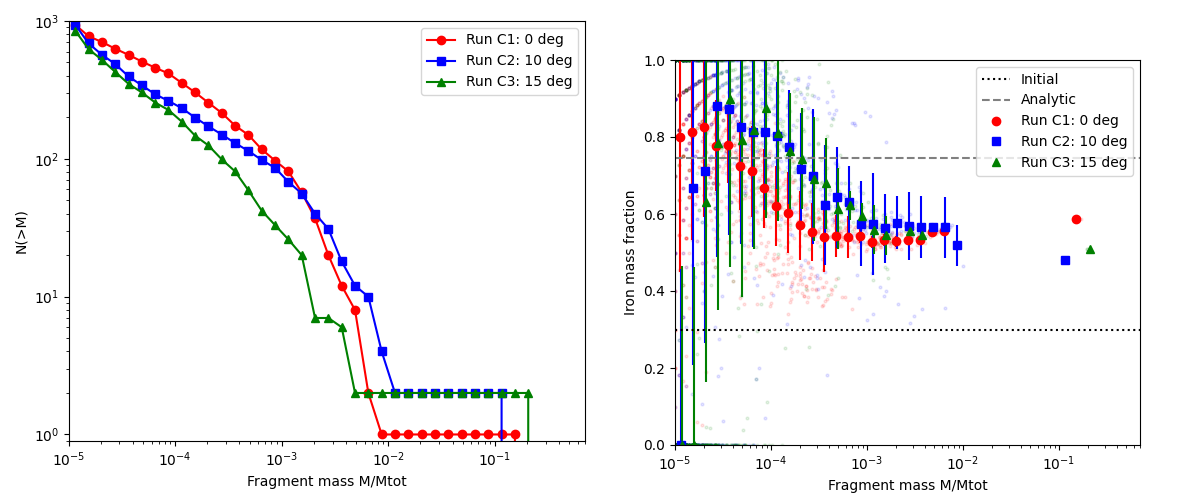}
    \caption{Fragments properties for the catastrophic disruption events. The target is 70\% of the rock mantle and 30\% of iron core mass and the impactor is identical to the target. The impact velocity is 1 km s$^{-1}$. The material strength of both the target and impactor is ignored. The Run numbers are listed in Table~\ref{tab:catastro}. left: Cumulative number distribution for a target with an iron core accounting for 30\% of the total mass. The different colors represent different impact angles: red, blue and green indicate impacts of $0^\circ$ (head-on), $10^\circ$, and $15^\circ$, respectively. right: The iron mass fraction for each remnant mass. The semi-transparent small points indicate the raw data before bin averaging. The dotted line indicates the pre‑impact iron mass fraction, and the dashed line shows the analytical result from Eq.~\ref{rock_prediction} described in section~\ref{dis:cd}.}
    \label{fig:I3R7}
\end{figure}

The results are summarized in Table~\ref{tab:catastro}.
Figure~\ref{fig:I3R7}a shows the mass and number distribution for the reaccumulated fragments from the target with 30\% of an iron core.
We compare three impact velocity and angle conditions: 1 km s$^{-1}$ with angles of 0$^\circ$ (head-on), $10^\circ$, and $15^\circ$.
In the head-on cases, both iron cores are fully disrupted, and the resulting iron-rich debris from the target and the impactor reaccretes into larger fragments.
On the other hand, the oblique impact cases, $10^\circ$ and $15^\circ$, result in a hit-and-run feature.
In our simulations, the outer rocky mantle is ejected at velocities well above the escape velocity and is dispersed over a wide solid angle.
As a result, the region near the impact point is rapidly depleted of rocky material, leaving predominantly iron-rich debris originating from the disrupted cores.
The rocky ejecta are typically very small (fewer than 10 SPH particles) and therefore fall below the resolution of our clump-finding (friend-of-friend) algorithm.
Within the short SPH timescale, these high-velocity rocky fragments do not interact again with the iron-rich debris within days, so the reaccreted fragments tend to retain similar iron-to-rock ratios.
Because our simulations do not include long-term orbital evolution, we cannot assess whether rocky ejecta might reaccrete onto iron-rich fragments at later times.
Tracking such late-time interactions would require N-body calculations over timescales of years, which is beyond the scope of this study.
Here we focus on the initial generation of iron-rich fragments immediately after the disruption.

On the other hand, in the case of an oblique impact, the rock mantle is lost, and the iron core cannot eject efficiently.
Small amounts of ejecta materials are reaccumulated after the impact.
Figure~\ref{fig:I3R7}b shows the relationship between the remnant masses and their iron mass fraction.
Our simulation indicated that the head-on impact produced many iron-rich bodies whose iron mass fractions were over 60\%, while most of the escaped rock material did not form clumps.

\begin{figure}
    \centering
    \includegraphics[bb=0 0 900 375,width=\linewidth]{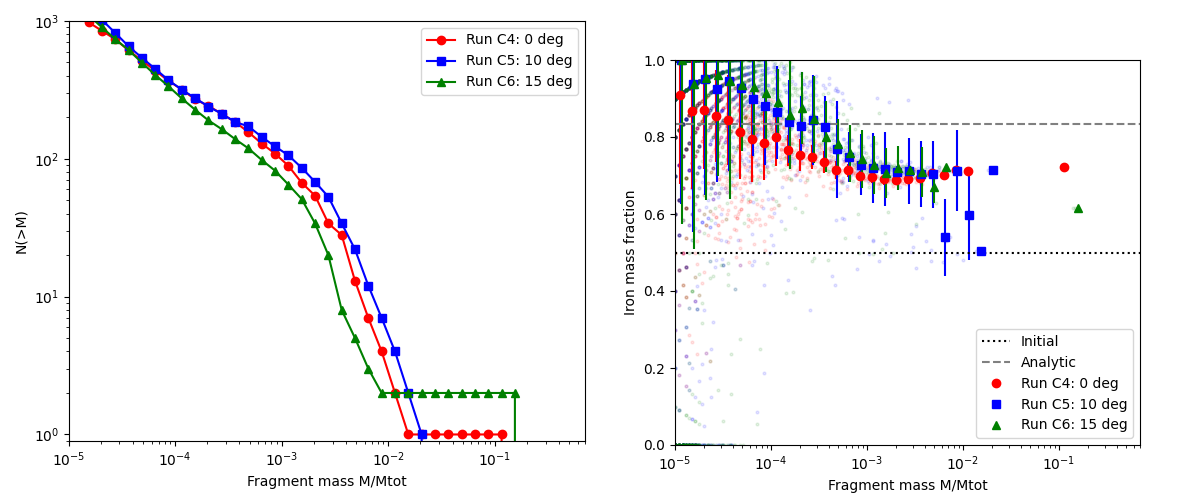}
    \caption{Fragments properties for the catastrophic disruption events. Identical to Figure~\ref{fig:I3R7} except that the target is 50\% of the rock mantle and 50\% of iron core mass and the impactor is identical to the target. The impact velocity is 1 km s$^{-1}$. The material strength of both the target and impactor is ignored. The Run numbers are listed in Table~\ref{tab:catastro}. The different colors represent different impact angles: red, blue and green indicate impacts of $0^\circ$ (head-on), $10^\circ$, and $15^\circ$, respectively.}
    \label{fig:I5R5}
\end{figure}

\begin{figure}
    \centering
    \includegraphics[bb=0 0 900 375,width=\linewidth]{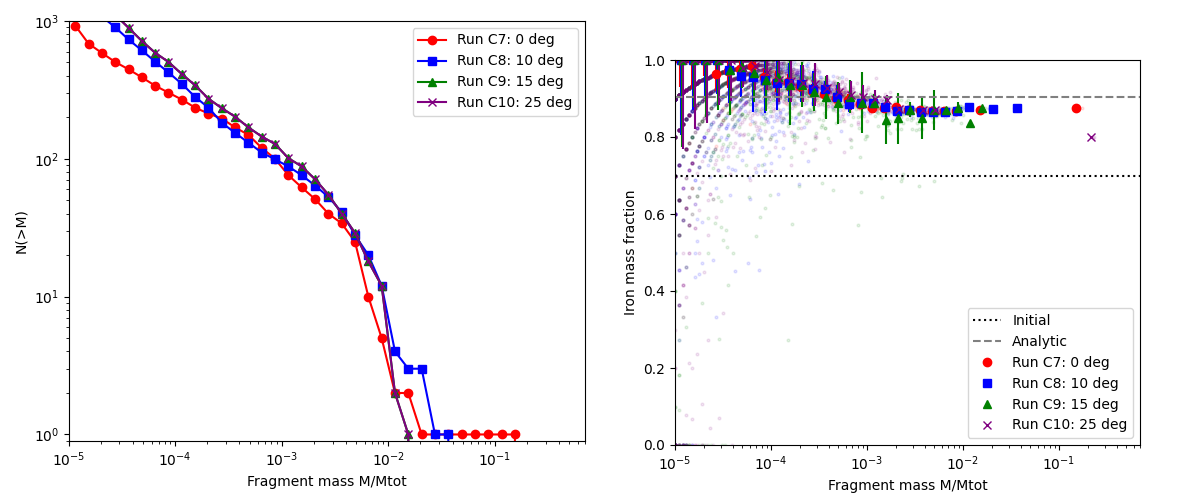}
    \caption{Fragments properties for the catastrophic disruption events. Identical to Figure~\ref{fig:I3R7} except that that the target is 30\% of the rock mantle and 70\% of iron core mass and the impactor is identical to the target. The impact velocity is 1 km s$^{-1}$. The material strength of both the target and impactor is ignored. The Run numbers are listed in Table~\ref{tab:catastro}. The different colors represent different impact angles: red, blue, green and purple indicate impacts of $0^\circ$ (head-on), $10^\circ$, $15^\circ$, and $25^\circ$ respectively.
    \label{fig:I7R3}}
\end{figure}

We also examined how different initial iron-mass fractions influence the resulting iron-mass fraction after catastrophic disruption.
Figures \ref{fig:I5R5} and \ref{fig:I7R3} show targets with iron cores accounting for 50\% and 70\% of their total mass, respectively.
The largest and second-largest fragment masses for head-on collisions are almost unchanged compared to the target with an iron core of 30\% of the total mass (Figure \ref{fig:I3R7}).
The size distributions for oblique impacts are also similar.
The iron mass fraction of fragments exhibits a different behavior.
As the iron core mass increases, the mass of the iron core fragments also increases.

\subsection{The effect of the material strength} \label{dis:mat-chk}
\begin{figure}
    \centering
    \includegraphics[bb=0 0 900 375,width=\linewidth]{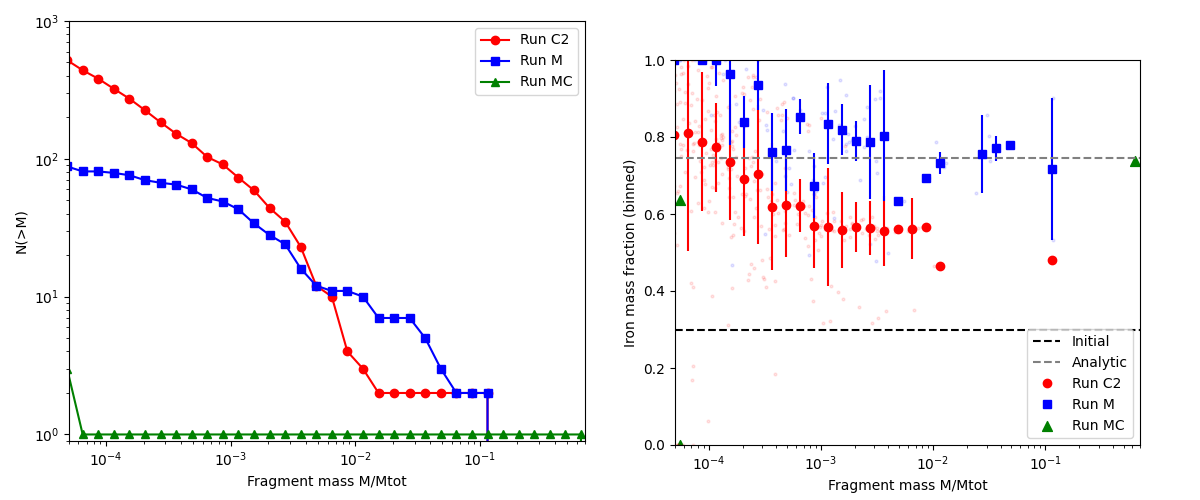}
    \caption{
    Fragment properties for the catastrophic-disruption events. The target has a rock‑mantle mass fraction of 70\% and an iron-core mass fraction of 30\%, and the impactor is identical to the target. The impact velocity is 1 km s$^{-1}$. Run M denotes the case in which only the mantle strength is included, and Run MC denotes the case in which both mantle and core strengths are included. The run numbers are listed in Table \ref{tab:catastro-dis}. Left panel: Cumulative number distribution for a target with an iron-core mass fraction of 30\%. The different colors represent different material models: red indicates no material strength, blue indicates mantle-only strength (Run M), and green indicates both mantle and core strengths (Run MC). Right panel: Iron mass fraction as a function of remnant mass. The semi-transparent small points indicate the raw data before bin averaging. The dotted line indicates the pre‑impact iron mass fraction, and the dashed line shows the analytical result from Eq.\ref{dis:cd}.
}
    \label{fig:RS}
\end{figure}

In this subsection, we present the results obtained when the material strength of both the mantle and the core is taken into account.
The calculations were performed using the same baseline conditions as Run C2. The target body has a mass of $10^{23}$ g, with a mantle mass fraction of 70~\% and a core mass fraction of 30~\%.
The impactor is identical to the target. The impact velocity is set to 1~km/s, and the impact angle is $10^\circ$.
The strength models employed for the rock mantle (Eqs. \ref{RYrock}, \ref{RYint}, \ref{RYdam}, \ref{RYtherm}) and for the iron core (Eq. \ref{JCmodel}) are summarized in Tables \ref{tab:mat-rock} and \ref{tab:mat-iron}, respectively.
We compare two cases: one in which only the mantle strength is included (Run M), and another in which both mantle and core strengths are included (Run MS).
Numerical results are listed in Table~\ref{tab:catastro-dis}.
Figure~\ref{fig:RS} shows the results.
When the rock mantle strength is considered, the mass of the largest remnant does not differ significantly from the strength-free case, but the mass distribution of smaller fragments changes.
Including rock mantle strength increases the number of fragments with masses of $10^{-2}- 10^{-1}M_{\mathrm{tot}}$, while reducing the number of fragments smaller than $10^{-2}M_{\mathrm{tot}}$.
This is likely because the frictional resistance of the mantle material suppresses fragmentation into very small pieces.

Furthermore, when iron core strength is included, catastrophic disruption does not occur under the same impact conditions as Run C2.
Instead, the two iron cores merge after the impact, while the rock mantle is stripped away.
The core strength increases the catastrophic disruption threshold, requiring much higher energy for catastrophic disruption.
In this study, the core is modeled using Johnson-Cook strength model informed by Fe-Ni impact experiments, and the assumption that the core behaves as a ductile material has a significant effect.
Because the rock mantle is brittle material, it undergoes immediate failure upon impact, resulting in fully damaged rock.
Consequently, the mantle strength after the impact is governed by the damaged rock strength.
In contrast, the ductile iron core does not experience a rapid strength drop due to damage by impact, and thus the influence of material strength becomes more pronounced.
However, 
our simulation ignore the failure of the iron core.
If an appropriate failure model for iron core were implemented, the outcome of catastrophic disruption could differ substantially.
This suggests that developing accurate failure models for high strain-rate environments is essential for future studies.

\section{Discussion} \label{discussion}

\subsection{Reaccumulated iron-rich asteroids generated by catastrophic disruption} \label{dis:cd}
\begin{figure}
    \centering
    \includegraphics[bb=0 0 900 375,width=\linewidth]{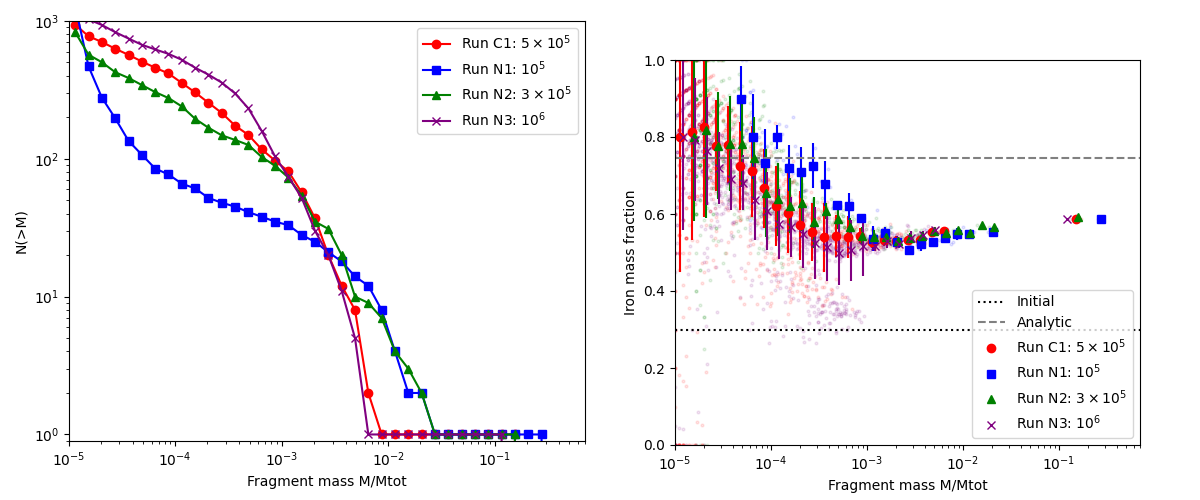}
    \caption{Fragments properties for the catastrophic disruption events. The target is 70\% of the rock mantle and 30\% of iron core mass and the impactor is identical to the target. The impact velocity is 1 km s$^{-1}$ with head-on collision. The material strength of both the target and impactor is ignored. The Run numbers are listed in Table~\ref{tab:catastro-N}. left: Cumulative number distribution for a target with an iron core accounting for 30\% of the total mass. The different colors represent different the number of particles: red, blue, green, and purple indicate $5\times10^5, 10^5, 3\times10^5$, and $10^6$, respectively. right: The iron mass fraction for each remnant mass. The semi-transparent small points indicate the raw data before bin averaging. The dotted line indicates the pre-impact iron mass fraction, and the dashed line shows the analytical result from Eq.~\ref{rock_prediction} described in section~\ref{dis:cd}.}
    \label{fig:D00chk}
\end{figure}

During a catastrophic disruption, the impact generates compression along the impact direction, while the material expands in the perpendicular plane.
This large deformation causes the disrupted body to spread into a flattened, sheet-like structure around the impact point.
Inside the sheet-like structure, small fragments are reaccreted due to their self-gravity.
The catastrophic disruption and re-accretion event requires a high-resolution calculation since the impact generated small fragments.
To verify the validity of the simulations, we compared the results for total particle numbers of $2\times10^5$, $1\times10^6$, and $2\times10^6$.
The iron mass fraction was set to 30~\% for impact angles of $0^\circ$ and $15^\circ$.
The results are summarized in Table~\ref{tab:catastro-N}, and Figure~\ref{fig:D00chk} shows the cumulative mass distribution and iron mass ratios of the fragments. 
The mass of the largest remnant remained consistent across different resolutions. The iron mass ratios of the fragments were also consistent across the different resolutions. 
The mass of the second-largest fragments converged for $10^6$ and $2\times10^6$ particles. The resolution dependence showed the same trend as the $15^\circ$ case. Thus, we adopted $10^6$ particles when considering the reaccretion results.

\begin{figure}
    \centering
    \includegraphics[bb=0 0 900 375,width=\linewidth]{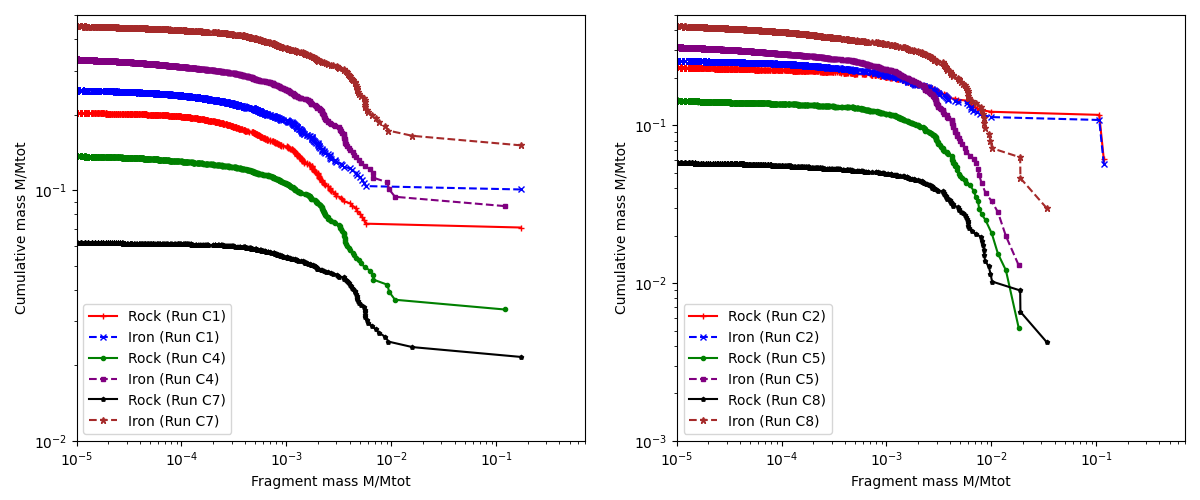} 
    \caption{Cumulative mass of the rock and iron components as a function of the total fragment mass. The left panel shows the head-on collision at 1 km s$^{-1}$, and the right panel shows the 10 degree impact at 1 km s$^{-1}$. Solid lines indicate the rock component, and dashed lines indicate the iron component. Red/blue, green/purple, and black/brown curves correspond to targets with initial iron mass fractions of 30\%, 50\%, and 70\%, respectively. The Run numbers shown in the legend correspond to the simulation settings listed in Table~\ref{tab:catastro}.}
    \label{fig:rock_iron_dis}
\end{figure}

To examine how the mass distribution of fragments depends on composition, we compared the cumulative mass distributions of rock and iron fragments for several catastrophic disruption cases.
Figure~\ref{fig:rock_iron_dis} shows the cumulative mass of each component as a function of the total fragment mass.
We analyzed head-on and $10^\circ$ at velocity 1 km/s impacts with different initial iron mass fractions.
In all cases, the cumulative distributions of rock and iron show nearly identical power-law slopes; the only difference is the total amount of each component, which shifts the curves vertically.
This indicates that, once catastrophic disruption occurs, the fragment mass distribution becomes largely independent of composition, impact angle, or initial core fraction.
The results suggest that the fragments produced in catastrophic disruptions tend to share an approximately uniform iron-rock ratio across a wide range of fragment masses, implying that reaccretion proceeds among fragments with similar compositions.

\begin{figure}
    \centering
    \includegraphics[bb=0 0 545 230,width=\linewidth]{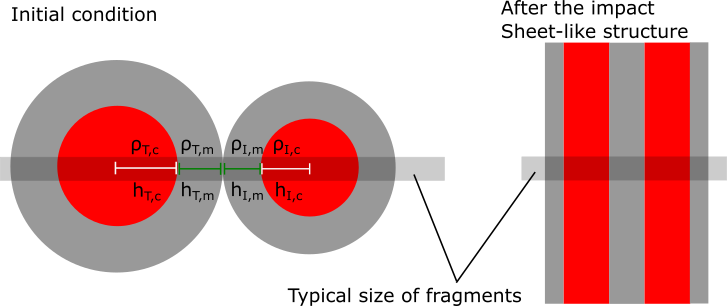}
    \caption{Schematic illustration of a catastrophic impact between differentiated bodies and the formation of an impact-generated sheet-like structure. The initial configuration consists of an iron core (target: density $\rho_\mathrm{T,c}$, thickness $h_\mathrm{T,c}$; impactor: density $\rho_\mathrm{I,c}$, thickness $h_\mathrm{I,c}$; shown in red) surrounded by a rocky mantle (target: density $\rho _\mathrm{T,m}$, thickness $h_\mathrm{T,m}$; impactor: density $\rho _\mathrm{I,m}$, thickness $h_\mathrm{I,m}$; shown in grey). After the collision, core and mantle materials are stretched into a sheet-like structure, from which fragments subsequently form by self-gravity. The shaded region indicates the typical size scale of a fragment, with characteristic radius $l$, used to estimate the pre-impact iron mass fraction of the material that reaccumulates into each fragment.}
    \label{fig:scmehatic}
\end{figure}

In catastrophic disruptions, the fragment population follows a nearly composition-independent cumulative mass distribution, allowing the iron mass fraction of fragments to be approximated by the bulk composition of the sheet-like structure formed immediately after impact.
Note that reaccumulated asteroids have similar iron mass fractions. 
We consider a scenario in which two differentiated asteroids collide and become compressed into a sheet-like structure.
Fragments are then formed from parts of this sheet-like structure due to its self-gravity. The iron mass fraction of each fragment is determined by the reaccumulation of specific regions indicated in Figure~\ref{fig:scmehatic}.
To estimate the iron mass ratio of these regions, we approximate their composition based on the pre-impact configuration.
We assume that the fragment's iron mass fraction corresponds to that of the shaded region in the figure, immediately prior to the collision.
We set $\rho_{\mathrm{T,c}}$ and $\rho_{\mathrm{T,m}}$ as the densities of the target’s iron core and rock mantle, respectively, and $h_{\mathrm{T,c}}$ and $h_{\mathrm{T,m}}$ as their corresponding thicknesses. 
Similarly, we set $\rho_{\mathrm{I,c}}$ and $\rho_{\mathrm{I,m}}$ as the densities of the impactor’s iron core and rock mantle, respectively, and $h_{\mathrm{I,c}}$ and $h_{\mathrm{I,m}}$ as their corresponding thicknesses. 
The typical fragment size is assumed to be $l$.
A predicted iron mass fraction $X_c^{\mathrm{pre}}$ of the fragment is then estimated using the following equation:
\begin{equation}
    X_c^{\mathrm{pre}} = \frac{(\rho_\mathrm{T,c} h_\mathrm{T,c} + \rho_\mathrm{I,c} h_\mathrm{I,c}) \pi l^2}{(\rho_\mathrm{T,c} h_\mathrm{T,c} + \rho_\mathrm{T,m} h_\mathrm{T,m} + \rho_\mathrm{I,c} h_\mathrm{I,c} + \rho_\mathrm{I,m} h_\mathrm{I,m}) \pi l^2} = \frac{1}{\frac{\rho_\mathrm{T,m} h_\mathrm{T,m} + \rho_\mathrm{I,m} h_\mathrm{I,m}}{\rho_\mathrm{T,c} h_\mathrm{T,c} + \rho_\mathrm{I,c} h_\mathrm{I,c}}+1}. \label{rock_prediction1}
\end{equation}
When the impactor is identical to the target, we set $\rho_\mathrm{T,m}=\rho_\mathrm{I,m}=\rho_\mathrm{m}$, $\rho_\mathrm{T,c}=\rho_\mathrm{I,c}=\rho_\mathrm{c}$,
$h_\mathrm{T,m}=h_\mathrm{I,m}=h_\mathrm{m}$, and $h_\mathrm{T,c}=h_\mathrm{I,c}=h_\mathrm{c}$.
These substitutions yield
\begin{equation}
    X_c^{\mathrm{pre}} =  \frac{1}{\frac{\rho_m h_m}{\rho_c h_c}+1}. \label{rock_prediction}
\end{equation}
We show the analytical prediction of $X_c^{\mathrm{pre}}$ in Figures~\ref{fig:I3R7}, \ref{fig:I5R5}, and \ref{fig:I7R3}.
Our results imply that the reaccumulation process after the impact of the catastrophic disruption event can generate iron-rich asteroids.
The sheet-structure model provides a first-order estimate of the iron mass fraction; however, the sheet undergoes viscous spreading after the impact, which reduces the local iron concentration.
This effect explains why the SPH-derived iron fractions are systematically lower than the analytical prediction. The rocky ejecta are also highly fragmented, and most rocky clumps contain fewer than 10 particles, falling below the resolution of our clump-finding algorithm.
The analytical model reproduces the composition of small fragments more accurately because these fragments originate from regions slightly offset from the impact point, where shock compression and subsequent spreading are weaker.
In contrast, larger fragments form closer to the impact point, where strong shock compression enhances sheet spreading and leads to deviations from the analytical estimate.
At higher resolution simulations, shock-induced deformation is captured more accurately even away from the impact point, which increases the discrepancy between the analytical model and the numerical results.
Our analysis focuses on the immediate post-impact structure.
Long-term viscous evolution and possible reaccretion of dispersed rocky material would require N-body simulations over much longer timescales and are left for future work.
In our simulations, fragments form from the impact-generated sheet-like structure, leading to clump formation.
Each clump typically reflects a characteristic size scale determined by self-gravity and local velocity dispersion.
These clumps emerge from the compressed sheet-like structure shown in Figure~\ref{fig:scmehatic}, and consequently share similar compositional properties, particularly in terms of iron mass fraction. Since they originate from neighboring regions within the sheet, the compositions among clumps remain comparable.
Once formed, clumps undergo gravitational interactions with nearby fragments and other clumps, gradually merging into larger fragments.
This hierarchical reaccumulation governs the final size distribution of fragments and preserves the compositional features of the configuration after the catastrophic impact.
The differentiated asteroids should have lost their rock mantle preferentially to make the iron-rich asteroids. The catastrophic disruption occurs when the impact energy exceeds $Q_R > 1.25\times 10^5$ J kg$^{-1}$, equivalent to when the $10^{20}$ kg target collides with the same mass impactor by 1 km s$^{-1}$ with a small impact angle.
Since the orbital velocity at the asteroid belt is 5 km s$^{-1}$ \citep[e.g.,][]{Bottke1994},
a Vesta-like planet may have experienced a catastrophic disruption impact event when another Vesta-like planet with a large eccentricity collides with a small impact angle.
After the catastrophic collision, iron-rich asteroids may have formed by reaccumulation of fragments, while most of the rock materials have been ejected. 
Thus, catastrophic disruption may form iron-rich asteroids, some of which may be present today in the M/X-type taxonomical group.

\subsection{Hit-and-run collision}
In this subsection, we discuss the hit-and-run impact case. Runs C10, C11, and C12 in Table \ref{tab:catastro} show the hit-and-run results.
In the hit-and-run case, the largest and second-largest fragments are almost the same as the pre-impact target and impactor because only small pieces
of the surface layers of the asteroids are ejected by the collision. 
Such hit-and-run cases eject the rock mantle of asteroids, while the ejection mass of iron is negligible. 
We find that the hit-and-run slightly increases the iron mass fraction of the asteroid.
However, the hit-and-run causes a smaller increase in the iron mass fraction compared to the catastrophic disruption.
Thus, a catastrophic collision can produce iron-rich asteroids or fragments more efficiently than a hit-and-run collision.

\subsection{Model dependence of EOS, artificial viscosity, and kernel function} \label{dis:valid}
In this subsection, we describe the effects of differences in simulation settings.
We adopted the Tillotson equation of state (EOS). In addition, the SPH simulations employed the von Neumann-Richtmeyer artificial viscosity and a Gaussian kernel function.
Here, following the approach in Section~\ref{dis:mat-chk}, we examine the influence of the EOS, artificial viscosity, and kernel function using Run M as the reference case.
The choice of EOS can potentially have a significant impact.
The Tillotson EOS used in this study does not compute temperature and therefore cannot correctly evaluate thermal pressure. To investigate the effects of heating in the outer rocky mantle during impacts, we performed additional simulations in which the mantle EOS was replaced with ANEOS basalt \citep{Pierazzo2005}.

\begin{figure}[htbp]
    \centering
    \includegraphics[bb=0 0 900 375,width=\linewidth]{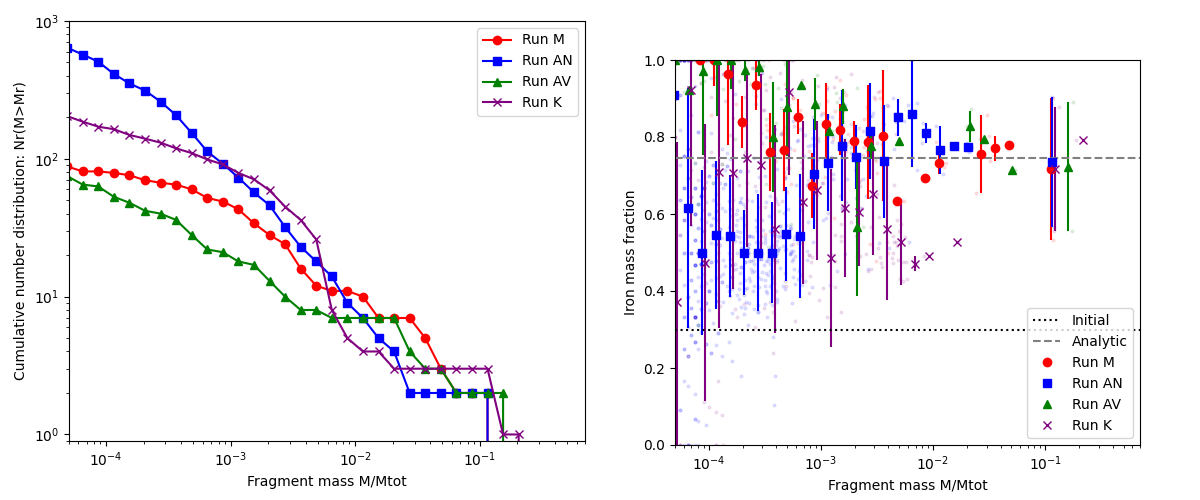}
    \caption{
    Fragment properties for the catastrophic-disruption events. The target has a rock-mantle mass fraction of 70\% and an iron-core mass fraction of 30\%, and the impactor is identical to the target. The impact velocity is 1 km s$^{-1}$. All simulations include mantle material strength but neglect core strength. Four SPH models are compared: Run M (nominal mantle-strength model), Run AN (ANEOS basalt), Run AV (artificial viscosity using Eq.\ref{pairav}), and Run K (kernel function using Eq.\ref{ker-S}). Note that all simulations include rock mantle material strength but neglect iron core material strength. The run numbers are listed in Table~\ref{tab:catastro-dis}. Left panel: Cumulative number distribution for a target with an iron-core mass fraction of 30\%. The colors represent different SPH models: red (Run M), blue (Run AN), green (Run AV), and purple (Run K). Right panel: Iron mass fraction as a function of remnant mass. Semi-transparent small points indicate the raw data before bin averaging. The dotted line shows the pre-impact iron mass fraction, and the dashed line shows the analytical result from Eq.\ref{dis:cd}.}
    \label{fig:eos}
\end{figure}

Figure~\ref{fig:eos} shows the results obtained using ANEOS for the rock mantle. 
The results indicate that, when ANEOS is used, the post-impact mass of the largest remnant decreases to roughly half of that obtained with the Tillotson EOS. This is likely because heating during the impact increases the contribution of thermal pressure, making it easier for the rocky mantle to be stripped away. Consequently, the resulting fragments tend to be smaller in size. Therefore, when discussing processes such as mantle vaporization or the behavior of gas components in detail, it is necessary to employ an EOS such as ANEOS, EOS for magma ocean \citep{Hosono2019,Hosono2022} that can accurately treat thermal effects.

Figure~\ref{fig:eos} shows the resulting fragment size distributions for different artificial viscosity models and kernel functions.
The artificial viscosity models and kernel functions are described in section~\ref{model:av-W}.
Although the overall trends are similar, artificial viscosity and the kernel function introduce measurable differences in the low-mass fragment distribution, while their effects on the largest remnant remain small. The differences arising from artificial viscosity likely reflect the enhanced shear viscosity between particles, which modifies the velocity dispersion among small fragments and consequently alters their tendency to merge. Because the Balsara switch was not implemented in this study, shear-related viscous effects become more pronounced in the small-fragment regime. For the kernel function, the differences can be attributed to the tendency for clumping. Compared with the Gaussian kernel, the spline kernel is more susceptible to pairing instability, which promotes the aggregation of small fragments and leads to the formation of larger fragments \citep[e.g.,][]{Dehnen2012}.

\subsection{Possibility of differentiation after the reaccretion}
Whether a reaccumulated fragments undergoes differentiation depends not only on its mass but also on the formation time of the pre-impact body. 
This is because the accretion time after CAI formation is a critical parameter for understanding differentiation \citep[e.g.,][]{Taylor1993}.
The reaccreted fragments have masses of $\le 10^{22}$ g, and bodies smaller than ~100 km may or may not differentiate depending on the timing of the collision.
The conditions for differentiation are complex and depend not only on body mass but also on the thermal evolution of each body’s magma ocean.
Among these factors, the accretion time after CAI formation plays a particularly important role
\citep{Moskovitz2011,Neumann2012,Lichtenberg2019,Lichtenberg2023,Monnereau2023}.
If the impact occurs onto a molten body that formed immediately after CAI formation, bodies larger than $\sim$20 km in radius corresponding to $\ge 10^{20}$ g in this study, are likely to have differentiated \citep[e.g.,][]{Monnereau2023}.
However, for formation times longer than 1.5 Myr after CAI formation, the radius required for differentiation becomes strongly dependent on the formation time, and even bodies larger than 20 km may fail to differentiate.
Although a detailed investigation of the thermal evolution of each body is beyond the scope of this work, the possibility that small reaccumulated bodies may or may not differentiate remains an important consideration.

\subsection{Implications for M-type asteroids: (16) Psyche and (22) Kalliope system}
In this study, we discuss the formation of iron-rich asteroids resulting from oblique impacts or the reaccumulation of fragments generated by catastrophic collisions.
Our results have useful implications for the M-type asteroid 16 Psyche, which has a mean diameter of 220 km and is expected to have a porosity that includes all scales from grains to rubble-pile fractures and cracks due to its mean density \citep[][and references therein]{Elkins-Tanton2020}.
The NASA Psyche mission is expected to improve our understanding of the surface conditions of iron-rich M-type asteroids.
When an iron-rich asteroid is formed by oblique impact, the asteroid is expected to have a layered structure with an iron core surrounded by a rock mantle because only the ejected rock mantle can reaccumulate on the asteroid's surface (see Figure \ref{fig:impact_example_small}).
On the other hand, an iron-rich asteroid that has experienced reaccumulation of fragments generated by a catastrophic disruption is formed by coagulated fragments composed of iron and rock.
Thus, the resultant asteroid is expected to be a rubble-pile object with iron components even on the surface (Figure \ref{fig:impact_example}). 
Therefore, our study suggests that 16 Psyche is formed by reaccumulating fragments generated by catastrophic disruption before the solidified the iron core if 16 Psyche is a rubble-pile object with iron fragments on its surface. 
The reaccumulation of compositionally differentiated fragments is expected to produce an asteroid with heterogeneous internal structure. As a result, its gravity field may show localized anomalies, potentially observable through high-resolution tracking of spacecraft trajectories during close encounters.

Our catastrophic disruption before the solidified the iron core scenario applies to the early epoch when the iron core of the parent body remained molten and material strength was negligible. \citet{Cambioni2026} also investigated collisions during the molten phase of planetesimals.
However, when the fragments are molten, reaccreted bodies equilibrate into a core mantle structure and do not form rubble-pile asteroids. In contrast, our results show that once the parent body has solidified, catastrophic disruption can produce metal-rich rubble-pile asteroids, particularly in low-angle collisions. Thus, our work reveals a new formation pathway that operates only after solidification, complementing the molten-phase scenarios explored in previous studies.

Whether a Psyche-mass fragment produced in our simulations represents a remnant core or a reaccreted rubble-pile depends on the thermal state of the parent body.
When the iron core is still molten and material strength is negligible, catastrophic disruption can fragment the core and produce Psyche-mass bodies through the reaccretion of iron-rich fragments.
In contrast, once the core has solidified, its strength inhibits catastrophic disruption, and Psyche-mass fragments are more likely to be the remnant cores of the parent bodies.
The iron mass fractions of Psyche-mass fragments in our simulations fall within the range inferred from Psyche’s bulk density, although the internal structure, a rubble-pile versus a remnant core, depends on whether the fragment forms by reaccretion or survival.
A full assessment of the long-term structural evolution requires thermal modeling beyond the scope of this study.

Our results show that rubble-pile-like fragments with masses and iron mass fractions comparable to Psyche can emerge as the largest members of iron-rock collisional families.
Such a formation pathway may also be relevant to the (22) Kalliope system, where a metal-rich primary, a satellite, and a mixed iron-rock collisional family are observed \citep{deKleer2024,Avdellidou2025}. 
The heterogeneous surface metal distribution predicted by our simulations is qualitatively consistent with ALMA observations of (16) Psyche and (22) Kalliope, which reveal spatial variations in surface metal content \citep{Cambioni2022,deKleer2024}.
These similarities suggest that catastrophic disruption and reaccretion of differentiated bodies can provide a unified framework for the formation of Psyche-like and Kalliope-like metal-rich asteroids, although a detailed exploration of specific impact configurations is left for future work.
We also recognize that future work incorporating higher-resolution SPH simulations together with N-body modeling of the orbital and collisional evolution of the fragments will be crucial for determining the size distribution and long-term behavior of the rubble-pile population.

Our finings are motivated by direct observations of metal-rich asteroid surfaces, including spectral and polarimetric evidence for Fe--Ni--rich regolith on M-type asteroids.
First, fragments would initially remain near the disrupted asteroid's orbital path. Unless external forces—such as gravitational scattering by nearby small bodies or effects like the Poynting-Robertson drag—remove or significantly perturb these fragments, bodies of similar mass and composition should remain in proximity for a long time.
Second, our simulations show that the fragment size distribution tends to be overestimated, resulting in the survival of only the largest fragment while smaller iron-rich bodies may be more dynamically unstable. These smaller fragments would be more susceptible to orbital modification or loss due to external forces, which may explain why fewer iron-rich asteroids are currently observed near the Psyche's orbit.
Consequently, if a concentration of iron-rich bodies were observed near Psyche's orbit, it could support the large-scale disruption hypothesis.
Conversely, the lack of such a population may point to alternative formation pathways or long-term dynamical evolution that dispersed or depleted the fragments.

\section{Conclusion} \label{conclusion}
In this study, we performed a series of impact simulations between differentiated bodies to investigate both mantle stripping and catastrophic disruption, as well as the subsequent reaccumulation of fragments.
Our results reveal that during catastrophic disruption, the rock mantle is preferentially ejected, and once approximately half of the mantle escapes, the iron core also becomes disrupted.
The catastrophic impact generates a sheet-like structure composed of stretched core and mantle materials, from which fragments subsequently form through self-gravity.
We also find that catastrophic disruption naturally produces numerous fragments with broadly similar iron to rock ratios.
The largest remnant is also formed by the reaccumulation of such compositionally similar fragments.

Even when the rock mantle is solidified and possesses material strength, catastrophic disruption still produces many small fragments, and their iron to rock ratios remain nearly uniform. However, including mantle strength reduces the number of small fragments compared to the strengthless case. In contrast, when the iron core is solidified, its strength suppresses catastrophic disruption, implying that the formation of numerous small fragments occurs primarily during epochs when the core remains molten. We also find that the fragment size distribution is sensitive not only to material strength but also to the choice of equation of state, highlighting the importance of both physical parameters in modeling differentiated-body collisions.

These findings have important implications for the surface and internal structure of rubble-pile asteroids.
If a metal-rich asteroid forms through oblique impact, it is expected to retain a layered structure with an iron core and rocky mantle.
In more head-on collisions, rubble-pile bodies can form as a result of catastrophic impact destruction followed by fragment reaccumulation into smaller bodies which exhibit iron-rich surface regions.

Our results provide insight into the formation of metal-rich asteroids such as (16) Psyche and the (22) Kalliope system.
The production and reaccumulation of numerous fragments with similar iron-rock ratios during catastrophic disruption may explain the origin of Psyche-like M-type asteroids and Kalliope-like systems with satellites.
The NASA Psyche mission will offer crucial observational constraints on these surface and internal characteristics, further advancing our understanding of iron-rich asteroids.


\appendix
\section{Fragment Size Distribution Analysis in Run C1} \label{app1}
\begin{figure}
    \centering
    \includegraphics[bb=0 0 480 346,width=\linewidth]{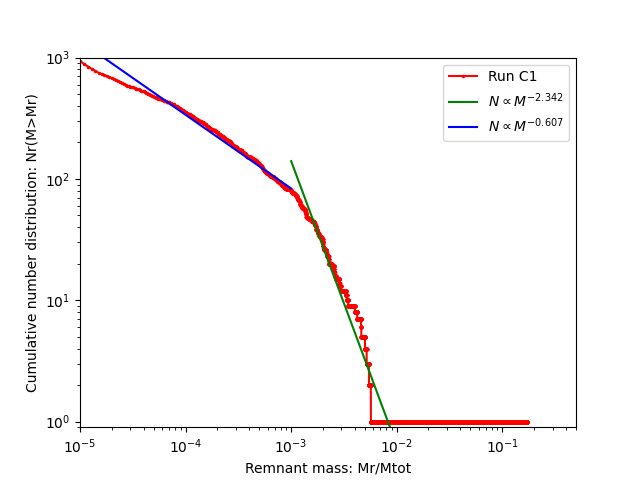}
    \caption{Cumulative number distribution of remnant mass for a head-on ($0^\circ$) collision involving a differentiated target with a 30\% iron core of the total mass. The red line shows the simulation results from Run C1. The blue and green lines represent power-law fits to the fragment mass ranges of $10^{-2}$--$10^{-3}$ and $10^{-3}$--$10^{-5}$, respectively.
}
    \label{fig:index}
\end{figure}

To characterize the cumulative number distribution of fragments produced by the impact event in Run C1, we fitted the simulation results using a power-law relationship of the form $N \propto M^b$ shown in Figure~\ref{fig:index}.
The fitted power-law indices were $b = -2.3$ for fragment masses between $10^{-2}$ and $10^{-3}$, and $b = -0.6$ for masses between $10^{-3}$ and $10^{-5}$.
These differences in slope suggest that reaccumulation plays a key role in modifying the fragment distribution.
In particular, fragments within the $10^{-2}$ to $10^{-3}$ mass range exhibit significant reaccumulation due to their relatively low velocity dispersion, allowing them to be gravitationally attracted to and merge with the largest surviving fragment. 
By contrast, fragments smaller than $10^{-3}$ are less influenced by gravitational interactions and tend to remain as isolated remnants.
This trend differs from the results of \citet{Kegerreis2025}, which investigated clump formation under tidal disruption using the ANEOS equation of state.
In their study, the power-law indices for small fragments ranged from $b = -0.4$ to $-0.5$.
We interpret this difference as arising not only from the equation of state employed, but also from the distinct mechanisms underlying clump generation, impact-induced reaccumulation in our study versus tidal fragmentation in theirs.

\section{Estimation of the catastrophic disruption energy}
This Appendix describes how we estimate the catastrophic disruption energies $Q_\textrm{D}^*$ and $Q_\textrm{MD}^*$.
We consider equal-mass collisions between a target and an impactor, each with mass $M_\textrm{T}=M_\textrm{I}=10^{23}$g. The initial iron‑core mass fraction of the bodies is varied (0\%, 30\%, 50\%, 70\%, and 100\%), and the impactor always has the same composition as the target. The total system mass is therefore $M_{\mathrm{tot}}=M_\textrm{T}+M_\textrm{I}$.
We denote the rocky‑mantle masses of the target and impactor as $M_\textrm{T}^M$ and $M_\textrm{I}^M$, and the total mantle mass as $M_{\mathrm{tot}}^M$.

The catastrophic disruption energy $Q_\textrm{D}^\ast$ is defined as the specific impact energy at which the gravitationally bound largest remnant has a mass $M_b/M_{\mathrm{tot}}=0.5$.
Because the iron core becomes increasingly difficult to disrupt as the core fraction increases, $Q_\textrm{D}^\ast$ grows with increasing initial iron-core mass fraction. This reflects the greater resistance of the core to fragmentation.

In this study, we also require a disruption threshold specific to the rocky mantle.
We therefore define $Q_\textrm{MD}^\ast$ as the specific impact energy at which the mantle mass of the largest remnant satisfies
$M_b^M/M_{\mathrm{tot}}^M=0.5$.
This quantity isolates the disruption of the mantle alone and is used in our scaling analysis of mantle stripping.
Unlike $Q_\textrm{D}^\ast$, the value of $Q_\textrm{MD}^\ast$ decreases with increasing iron-core mass fraction, because the total mantle mass becomes smaller and therefore easier to remove.

\begin{figure}
    \centering
    \includegraphics[bb=0 0 960 360,width=\linewidth]{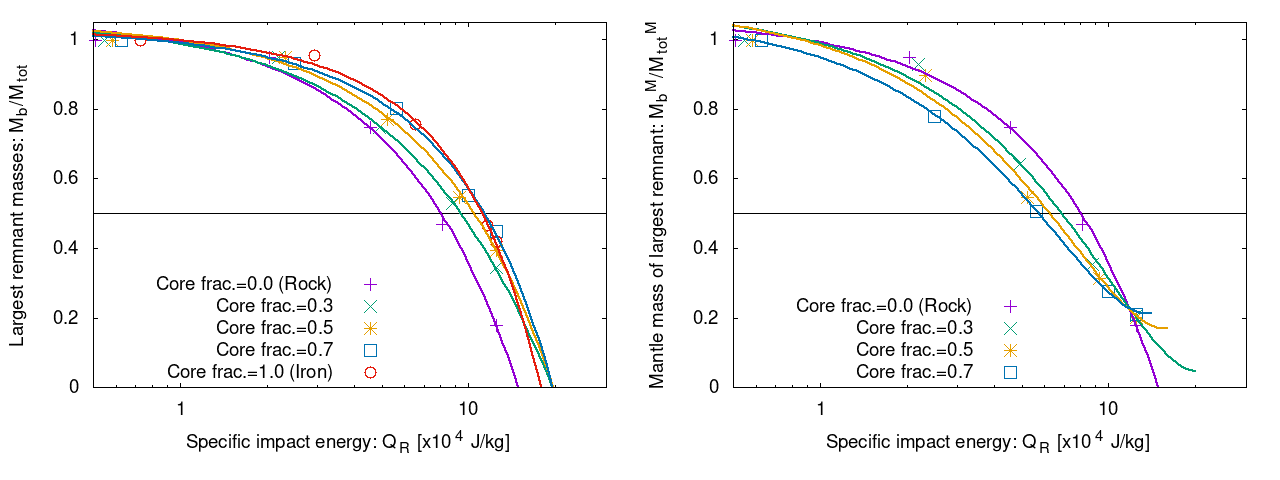}
    \caption{Largest-remnant mass $M_b/M_\textrm{tot}$ (left) and mantle mass of the largest remnant $M_b^M/M_\textrm{tot}^M$ (right) as functions of the specific impact energy $Q_R$. The target mass is $10^{23}$g, and the impactor has the same mass and composition as the target. The specific impact energy is varied by changing the impact velocity. Purple (plus), green (cross), orange (asterisk), blue (square), and red (circle) symbols correspond to initial iron-core mass fractions of 0\%, 30\%, 50\%, 70\%, and 100\%, respectively. For each composition, the simulation results are fitted with a quadratic function, and the values of $Q_R$ at which $M_b/M_\textrm{tot}=0.5$ and $M_b^M/M_\textrm{tot}^M=0.5$
are taken as $Q_\textrm{D}^{\ast}$ and $Q_\textrm{MD}^{\ast}$, respectively.}
    \label{fig:QDQMD}
\end{figure}

Figure~\ref{fig:QDQMD} shows the mass of the gravitationally bound largest remnant as a function of the specific impact energy for the equal-mass collisions described above.
For each composition, we fit the simulation results with a quadratic function using a least-squares method.
From the fitted curve, we determine the values of $Q_R$ at which
$M_b/M_{\mathrm{tot}}=0.5$ and $M_b^M/M_{\mathrm{tot}}^M=0.5$,
yielding $Q_\textrm{D}^\ast$ and $Q_\textrm{MD}^\ast$, respectively.
The resulting values are listed in Table~\ref{tab:qdstar}.

The values obtained here represent minimum estimates of the catastrophic disruption energies, because $Q_\textrm{D}^\ast$ is known to depend on both impact angle and impactor mass.
Oblique impacts or smaller impactors generally require larger specific energies to achieve catastrophic disruption.
Previous studies have also shown that $Q_\textrm{D}^\ast$ can depend not only on impact velocity but also on projectile size \citep{Benz1999,Jutzi2010}.
Although our simulations focus on equal-mass collisions, this dependence implies that the catastrophic disruption threshold may vary for different impactor to target mass ratios. A detailed exploration of these dependencies is beyond the scope of this study.

\section{Heating by the catastrophic impact}
In this appendix, we analyze the heating that occurs during catastrophic disruption.
To estimate the amount of heating, we computed the difference between the maximum internal energy reached after the impact and the initial internal energy of each SPH particle. Using the difference of the internal energy, we constructed cumulative mass distributions of impact-induced increases in internal energy.

To evaluate the role of material strength in impact heating, we compared three simulations: no strength (Run C2), mantle strength only (Run M), and both mantle and core strength (Run MC).
As a reference threshold for significant heating, we used the internal energy required for the onset of partial melting of basalt,
$E_{\mathrm{iv}}=4.72\times 10^{10}\ \mathrm{erg\ g^{-1}}$  \citep{Benz1999}.
To compute the cumulative mass distribution, we divided the internal energy increase into 50 logarithmically spaced bins spanning the full dynamic range between its minimum and maximum values.

\begin{figure}[htbp]
    \centering
    \includegraphics[bb=0 0 1200 450,width=\linewidth]{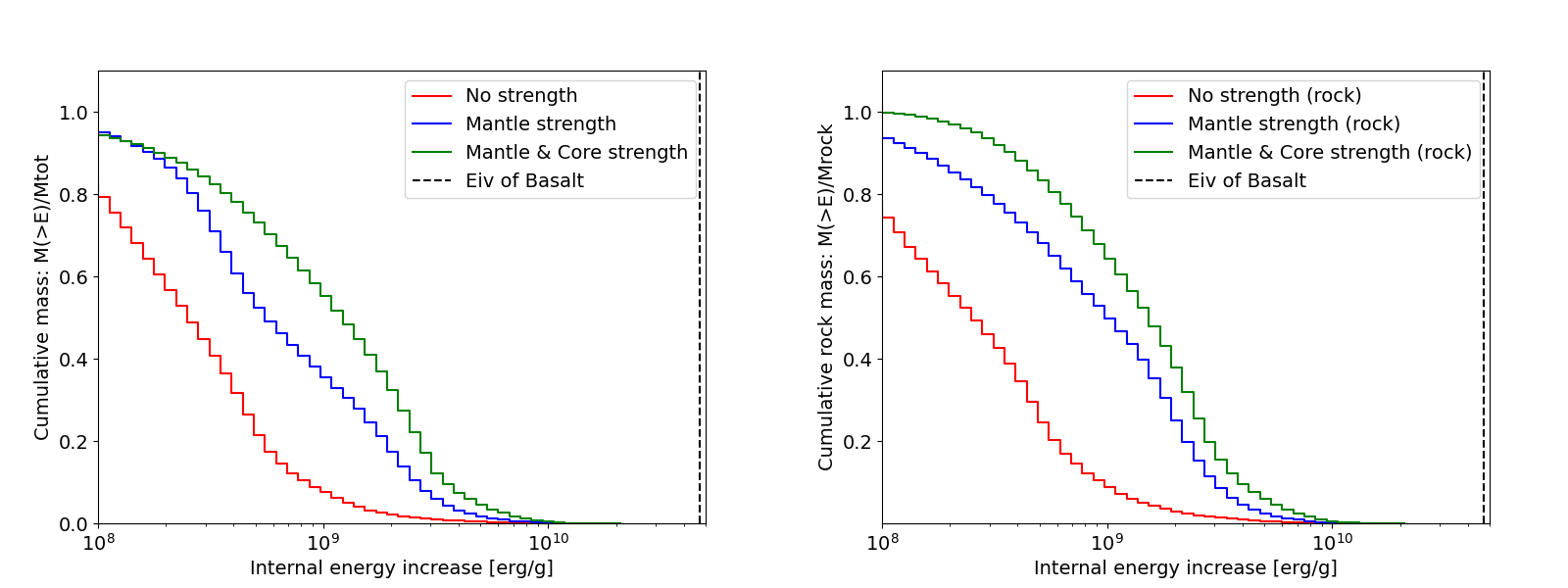}
    \caption{Internal energy increases for the catastrophic‑disruption events. The target has a rock-mantle mass fraction of 70\% and an iron-core mass fraction of 30\%, and the impactor is identical to the target. The impact velocity is 1 km s$^{-1}$. "No strength" denotes the case in which there is no material strength (Run C2), "Mantle strength" denotes the case in which only the mantle strength is included (Run M), and "Mantle \& Core strength" denotes the case in which both mantle and core strengths are included (Run MC). The run numbers are listed in Table \ref{tab:catastro-dis}. Left panel: Cumulative mass distribution of the internal energy increase. The cumulative mass is defined as the fraction of total mass (rock + iron) whose internal-energy increase exceeds a given value. The different colors represent different material models: red indicates no material strength, blue indicates mantle-only strength (Run M), and green indicates both mantle and core strengths (Run MC). The vertical black dash line represents the $E_\mathrm{iv}$ of the basalt \citep{Benz1999}. Right panel: Cumulative rock mass distribution of the internal energy increase. The cumulative mass is defined as the fraction of rock mass whose internal energy increase exceeds a given value.}
    \label{fig:engimp}
\end{figure}

Figure~\ref{fig:engimp} shows the cumulative mass distributions of internal-energy increase for the total mass (rock + iron) and for the rock component alone. In all simulations, the increase in internal energy remains well below the basalt partial-melting threshold. The fraction of material experiencing an internal-energy increase greater than $10^{10}\ \mathrm{erg\ g^{-1}}$ is less than 1\% of the total mass. These results are consistent with previous studies showing that heating is inefficient at low impact velocities for a non-porous body 
\citep[e.g.,][]{Keil1997,Davison2010}.

We also find that the magnitude of the internal-energy increase depends on the assumed material strength:
The no-strength case exhibits the smallest heating, followed by the mantle-strength case, and the mantle-plus-core-strength case shows the largest heating. This trend reflects the additional frictional dissipation introduced by material strength \citep[e.g.,][]{Kurosawa2018}.
A similar trend is observed when considering only the rock mantle component.


\begin{thebibliography}{00}

\bibitem[Alexander et al.(2022)]{Alexander2022} Alexander, A.~M., Marchi, S., Chocron, S., Walker, J.\ 2022.\ Benchmarking iSALE and CTH Shock Physics Codes to In Situ High-Velocity Impact Experiments Into Fe-Ni Targets.\ Earth and Space Science 9. doi:10.1029/2021EA00199210.1002/essoar.10507852.1
\bibitem[Asphaug and Reufer(2014)]{Asphaug2014} Asphaug, E., Reufer, A.\ 2014.\ Mercury and other iron-rich planetary bodies as relics of inefficient accretion.\ Nature Geoscience 7, 564–568. doi:10.1038/ngeo2189
\bibitem[Avdellidou et al.(2025)]{Avdellidou2025} Avdellidou, C., Bhat, U., Bujdoso, K., Delbo, M., Marsset, M., Vernazza, P.\ 2025.\ Kalliope sings rock and metal.\ Monthly Notices of the Royal Astronomical Society 539, 3534–3550. doi:10.1093/mnras/staf640
\bibitem[Benavidez et al.(2012)]{Benavidez2012} Benavidez, P.~G. and 7 colleagues 2012.\ A comparison between rubble-pile and monolithic targets in impact simulations: Application to asteroid satellites and family size distributions.\ Icarus 219, 57–76. doi:10.1016/j.icarus.2012.01.015
\bibitem[Benz et al.(1988)]{Benz1988} Benz, W., Slattery, W.~L., Cameron, A.~G.~W.\ 1988.\ Collisional stripping of Mercury's mantle.\ Icarus 74, 516–528. doi:10.1016/0019-1035(88)90118-2
\bibitem[Benz and Asphaug(1994)]{Benz1994} Benz, W., Asphaug, E.\ 1994.\ Impact Simulations with Fracture. I. Method and Tests.\ Icarus 107, 98–116. doi:10.1006/icar.1994.1009
\bibitem[Benz and Asphaug(1995)]{Benz1995} Benz, W., Asphaug, E.\ 1995.\ Simulations of brittle solids using smooth particle hydrodynamics.\ Computer Physics Communications 87, 253–265. doi:10.1016/0010-4655(94)00176-3
\bibitem[Benz and Asphaug(1999)]{Benz1999} Benz, W., Asphaug, E.\ 1999.\ Catastrophic Disruptions Revisited.\ Icarus 142, 5–20. doi:10.1006/icar.1999.6204
\bibitem[Benz et al.(2007)]{Benz2007} Benz, W., Anic, A., Horner, J., Whitby, J.~A.\ 2007.\ The Origin of Mercury.\ Space Science Reviews 132, 189–202. doi:10.1007/s11214-007-9284-1
\bibitem[Bottke et al.(1994)]{Bottke1994} Bottke, W.~F., Nolan, M.~C., Greenberg, R., Kolvoord, R.~A.\ 1994.\ Velocity Distributions among Colliding Asteroids.\ Icarus 107, 255–268. doi:10.1006/icar.1994.1021
\bibitem[Bottke et al.(2006)]{Bottke2006} Bottke, W.~F., Nesvorn{\'y}, D., Grimm, R.~E., Morbidelli, A., O'Brien, D.~P.\ 2006.\ Iron meteorites as remnants of planetesimals formed in the terrestrial planet region.\ Nature 439, 821–824. doi:10.1038/nature04536
\bibitem[Burbine et al.(2002)]{Burbine2002} Burbine, T.~H., McCoy, T.~J., Meibom, A., Gladman, B., Keil, K.\ 2002.\ Meteoritic Parent Bodies: Their Number and Identification.\ Asteroids III 653–667.
\bibitem[Bus and Binzel(2002)]{Bus2002} Bus, S.~J., Binzel, R.~P.\ 2002.\ Phase II of the Small Main-Belt Asteroid Spectroscopic Survey. A Feature-Based Taxonomy.\ Icarus 158, 146–177. doi:10.1006/icar.2002.6856
\bibitem[Cambioni et al.(2022)]{Cambioni2022} Cambioni, S., de Kleer, K., Shepard, M.\ 2022.\ The Heterogeneous Surface of Asteroid (16) Psyche.\ Journal of Geophysical Research (Planets) 127. doi:10.1029/2021JE007091
\bibitem[Cambioni et al.(2025)]{Cambioni2025} Cambioni, S. and 7 colleagues 2025.\ Can metal-rich worlds form by giant impacts?.\ Astronomy and Astrophysics 696. doi:10.1051/0004-6361/202450128
\bibitem[Cambioni et al.(2026)]{Cambioni2026} Cambioni, S. and 14 colleagues 2026.\ Formation of Asteroid (16) Psyche by a Giant Impact.\ Journal of Geophysical Research (Planets) 131. doi:10.1029/2025JE009317
\bibitem[Carter et al.(2018)]{Carter2018} Carter, P.~J., Leinhardt, Z.~M., Elliott, T., Stewart, S.~T., Walter, M.~J.\ 2018.\ Collisional stripping of planetary crusts.\ Earth and Planetary Science Letters 484, 276–286. doi:10.1016/j.epsl.2017.12.012
\bibitem[Chau et al.(2018)]{Chau2018} Chau, A., Reinhardt, C., Helled, R., Stadel, J.\ 2018.\ Forming Mercury by Giant Impacts.\ The Astrophysical Journal 865. doi:10.3847/1538-4357/aad8b0
\bibitem[Cloutis et al.(2010)]{Cloutis2010} Cloutis, E.~A., Hardersen, P.~S., Bish, D.~L., Bailey, D.~T., Gaffey, M.~J., Craig, M.~A.\ 2010.\ Reflectance spectra of iron meteorites: Implications for spectral identification of their parent bodies.\ Meteoritics and Planetary Science 45, 304–332. doi:10.1111/j.1945-5100.2010.01033.x
\bibitem[Collins et al.(2004)]{Collins2004} Collins, G.~S., Melosh, H.~J., Ivanov, B.~A.\ 2004.\ Modeling damage and deformation in impact simulations.\ Meteoritics and Planetary Science 39, 217–231. doi:10.1111/j.1945-5100.2004.tb00337.x
\bibitem[Davison et al.(2010)]{Davison2010} Davison, T.~M., Collins, G.~S., Ciesla, F.~J.\ 2010.\ Numerical modelling of heating in porous planetesimal collisions.\ Icarus 208, 468–481. doi:10.1016/j.icarus.2010.01.034
\bibitem[Dehnen and Aly(2012)]{Dehnen2012} Dehnen, W., Aly, H.\ 2012.\ Improving convergence in smoothed particle hydrodynamics simulations without pairing instability.\ Monthly Notices of the Royal Astronomical Society 425, 1068–1082. doi:10.1111/j.1365-2966.2012.21439.x
\bibitem[de Kleer et al.(2021)]{deKleer2021} de Kleer, K., Cambioni, S., Shepard, M.\ 2021.\ The Surface of (16) Psyche from Thermal Emission and Polarization Mapping.\ The Planetary Science Journal 2. doi:10.3847/PSJ/ac01ec
\bibitem[de Kleer et al.(2024)]{deKleer2024} de Kleer, K., Cambioni, S., Butler, B., Shepard, M.\ 2024.\ Surface Properties of the Kalliope{\textendash}Linus System from ALMA and VLA Data.\ The Planetary Science Journal 5. doi:10.3847/PSJ/ad7797
\bibitem[DeMeo et al.(2009)]{DeMeo2009} DeMeo, F.~E., Binzel, R.~P., Slivan, S.~M., Bus, S.~J.\ 2009.\ An extension of the Bus asteroid taxonomy into the near-infrared.\ Icarus 202, 160–180. doi:10.1016/j.icarus.2009.02.005
\bibitem[Dibb et al.(2023)]{Dibb2023} Dibb, S.~D., Bell, J.~F., Elkins-Tanton, L.~T., Williams, D.~A.\ 2023.\ Visible to Near-Infrared Reflectance Spectroscopy of Asteroid (16) Psyche: Implications for the Psyche Mission's Science Investigations.\ Earth and Space Science 10. doi:10.1029/2022EA002694
\bibitem[Dibb et al.(2024)]{Dibb2024} Dibb, S.~D. and 15 colleagues 2024.\ A Post-Launch Summary of the Science of NASA's Psyche Mission.\ AGU Advances 5. doi:10.1029/2023AV001077
\bibitem[Dodds et al.(2021)]{Dodds2021} Dodds, K.~H., Bryson, J.~F.~J., Neufeld, J.~A., Harrison, R.~J.\ 2021.\ The Thermal Evolution of Planetesimals During Accretion and Differentiation: Consequences for Dynamo Generation by Thermally Driven Convection.\ Journal of Geophysical Research (Planets) 126. doi:10.1029/2020JE00670410.1002/essoar.10504425.1
\bibitem[Dollfus et al.(1979)]{Dollfus1979} Dollfus, A., Mandeville, J.~C., Duseaux, M.\ 1979.\ The nature of the M-type asteroids from optical polarimetry.\ Icarus 37, 124–132. doi:10.1016/0019-1035(79)90120-9
\bibitem[Dou et al.(2024)]{Dou2024} Dou, J., Carter, P.~J., Leinhardt, Z.~M.\ 2024.\ Formation of super-Mercuries via giant impacts.\ Monthly Notices of the Royal Astronomical Society 529, 2577–2594. doi:10.1093/mnras/stae644
\bibitem[Durda et al.(2007)]{Durda2007} Durda, D.~D. and 6 colleagues 2007.\ Size-frequency distributions of fragments from SPH/ N-body simulations of asteroid impacts: Comparison with observed asteroid families.\ Icarus 186, 498–516. doi:10.1016/j.icarus.2006.09.013
\bibitem[Elkins-Tanton et al.(2011)]{Elkins-Tanton2011} Elkins-Tanton, L.~T., Weiss, B.~P., Zuber, M.~T.\ 2011.\ Chondrites as samples of differentiated planetesimals.\ Earth and Planetary Science Letters 305, 1–10. doi:10.1016/j.epsl.2011.03.010
\bibitem[Elkins-Tanton et al.(2020)]{Elkins-Tanton2020} Elkins-Tanton, L.~T. and 21 colleagues 2020.\ Observations, Meteorites, and Models: A Preflight Assessment of the Composition and Formation of (16) Psyche.\ Journal of Geophysical Research (Planets) 125. doi:10.1029/2019JE006296
\bibitem[Elkins-Tanton et al.(2022)]{Elkins-Tanton2022} Elkins-Tanton, L.~T. and 21 colleagues 2022.\ Distinguishing the Origin of Asteroid (16) Psyche.\ Space Science Reviews 218. doi:10.1007/s11214-022-00880-9
\bibitem[Emsenhuber et al.(2018)]{Emsenhuber2018} Emsenhuber, A., Jutzi, M., Benz, W.\ 2018.\ SPH calculations of Mars-scale collisions: The role of the equation of state, material rheologies, and numerical effects.\ Icarus 301, 247–257. doi:10.1016/j.icarus.2017.09.017
\bibitem[Emsenhuber et al.(2024)]{Emsenhuber2024} Emsenhuber, A. and 6 colleagues 2024.\ A New Database of Giant Impacts over a Wide Range of Masses and with Material Strength: A First Analysis of Outcomes.\ The Planetary Science Journal 5. doi:10.3847/PSJ/ad2178
\bibitem[Farinella et al.(1982)]{Farinella1982} Farinella, P., Paolicchi, P., Zappala, V.\ 1982.\ The asteroids as outcomes of catastrophic collisions.\ Icarus 52, 409–433. doi:10.1016/0019-1035(82)90003-3
\bibitem[Farnocchia et al.(2024)]{Farnocchia2024} Farnocchia, D., Fuentes-Mu{\~n}oz, O., Park, R.~S., Baer, J., Chesley, S.~R.\ 2024.\ Mass, Density, and Radius of Asteroid (16) Psyche from High-precision Astrometry.\ The Astronomical Journal 168. doi:10.3847/1538-3881/ad50ca
\bibitem[Franco et al.(2022)]{Franco2022} Franco, P., Izidoro, A., Winter, O.~C., Torres, K.~S., Amarante, A.\ 2022.\ Explaining mercury via a single giant impact is highly unlikely.\ Monthly Notices of the Royal Astronomical Society 515, 5576–5586. doi:10.1093/mnras/stac2183
\bibitem[Gabriel and Cambioni(2023)]{Gabriel2023} Gabriel, T.~S.~J., Cambioni, S.\ 2023.\ The Role of Giant Impacts in Planet Formation.\ Annual Review of Earth and Planetary Sciences 51, 671–695. doi:10.1146/annurev-earth-031621-055545
\bibitem[Genda et al.(2017)]{Genda2017} Genda, H., Fujita, T., Kobayashi, H., Tanaka, H., Suetsugu, R., Abe, Y.\ 2017.\ Impact erosion model for gravity-dominated planetesimals.\ Icarus 294, 234–246. doi:10.1016/j.icarus.2017.03.009
\bibitem[Grady and Kipp(1980)]{Grady1980} Grady, D.~E., Kipp, M.~E.\ 1980.\ Continuum modelling of explosive fracture in oil shale.\ International Journal of Rock Mechanics and Mining Sciences and Geomechanics Abstracts 17, 147–157. doi:10.1016/0148-9062(80)91361-3
\bibitem[Hardersen et al.(2005)]{Hardersen2005} Hardersen, P.~S., Gaffey, M.~J., Abell, P.~A.\ 2005.\ Near-IR spectral evidence for the presence of iron-poor orthopyroxenes on the surfaces of six M-type asteroids.\ Icarus 175, 141–158. doi:10.1016/j.icarus.2004.10.017
\bibitem[Hongbin and Xin(2005)]{Hongbin2005} Hongbin, J., Xin, D.\ 2005.\ On criterions for smoothed particle hydrodynamics kernels in stable field.\ Journal of Computational Physics 202, 699–709. doi:10.1016/j.jcp.2004.08.002
\bibitem[Hosono and Karato(2022)]{Hosono2022} Hosono, N., Karato, S.-. ichiro .\ 2022.\ The Influence of Equation of State on the Giant Impact Simulations.\ Journal of Geophysical Research (Planets) 127. doi:10.1029/2021JE006971
\bibitem[Hosono et al.(2019)]{Hosono2019} Hosono, N., Karato, S.-. ichiro ., Makino, J., Saitoh, T.~R.\ 2019.\ Terrestrial magma ocean origin of the Moon.\ Nature Geoscience 12, 418–423. doi:10.1038/s41561-019-0354-2
\bibitem[Hosono et al.(2016)]{Hosono2016} Hosono, N., Saitoh, T.~R., Makino, J.\ 2016.\ A Comparison of SPH Artificial Viscosities and Their Impact on the Keplerian Disk.\ The Astrophysical Journal Supplement Series 224. doi:10.3847/0067-0049/224/2/32
\bibitem[Hyodo et al.(2017)]{Hyodo2017} Hyodo, R., Charnoz, S., Ohtsuki, K., Genda, H.\ 2017.\ Ring formation around giant planets by tidal disruption of a single passing large Kuiper belt object.\ Icarus 282, 195–213. doi:10.1016/j.icarus.2016.09.012
\bibitem[Iwasawa et al.(2016)]{Iwasawa2016} Iwasawa, M., Tanikawa, A., Hosono, N., Nitadori, K., Muranushi, T., Makino, J.\ 2016.\ Implementation and performance of FDPS: a framework for developing parallel particle simulation codes.\ Publications of the Astronomical Society of Japan 68. doi:10.1093/pasj/psw053
\bibitem[Johnson and Cook (1983)]{Johnson1983} Johnson, G.R. and Cook, W.H.\ 1983.\ A Constitutive Model and Data for Metals Subjected to Large Strains, High Strain Rates, and High Temperatures. Proceedings 7th International Symposium on Ballistics, The Hague, 19-21 April 1983, 541-547.
\bibitem[Johnson et al.(2014)]{Johnson2014} Johnson, B.~C., Bowling, T.~J., Melosh, H.~J.\ 2014.\ Jetting during vertical impacts of spherical projectiles.\ Icarus 238, 13–22. doi:10.1016/j.icarus.2014.05.003
\bibitem[Jutzi et al.(2010)]{Jutzi2010} Jutzi, M., Michel, P., Benz, W., Richardson, D.~C.\ 2010.\ Fragment properties at the catastrophic disruption threshold: The effect of the parent body{\textquoteright}s internal structure.\ Icarus 207, 54–65. doi:10.1016/j.icarus.2009.11.016
\bibitem[Jutzi et al.(2013)]{Jutzi2013} Jutzi, M., Asphaug, E., Gillet, P., Barrat, J.-A., Benz, W.\ 2013.\ The structure of the asteroid 4 Vesta as revealed by models of planet-scale collisions.\ Nature 494, 207–210. doi:10.1038/nature11892
\bibitem[Jutzi(2015)]{Jutzi2015} Jutzi, M.\ 2015.\ SPH calculations of asteroid disruptions: The role of pressure dependent failure models.\ Planetary and Space Science 107, 3–9. doi:10.1016/j.pss.2014.09.012
\bibitem[Jutzi et al.(2019)]{Jutzi2019b} Jutzi, M., Michel, P., Richardson, D.~C.\ 2019.\ Fragment properties from large-scale asteroid collisions: I: Results from SPH/N-body simulations using porous parent bodies and improved material models.\ Icarus 317, 215–228. doi:10.1016/j.icarus.2018.08.006
\bibitem[Jutzi and Michel(2020)]{Jutzi2020} Jutzi, M., Michel, P.\ 2020.\ Collisional heating and compaction of small bodies: Constraints for their origin and evolution.\ Icarus 350. doi:10.1016/j.icarus.2020.113867
\bibitem[Kegerreis et al.(2020a)]{Kegerreis2020a} Kegerreis, J.~A., Eke, V.~R., Massey, R.~J., Teodoro, L.~F.~A.\ 2020.\ Atmospheric Erosion by Giant Impacts onto Terrestrial Planets.\ The Astrophysical Journal 897. doi:10.3847/1538-4357/ab9810
\bibitem[Kegerreis et al.(2020b)]{Kegerreis2020b} Kegerreis, J.~A., Eke, V.~R., Catling, D.~C., Massey, R.~J., Teodoro, L.~F.~A., Zahnle, K.~J.\ 2020.\ Atmospheric Erosion by Giant Impacts onto Terrestrial Planets: A Scaling Law for any Speed, Angle, Mass, and Density.\ The Astrophysical Journal 901. doi:10.3847/2041-8213/abb5fb
\bibitem[Kegerreis et al.(2025)]{Kegerreis2025} Kegerreis, J.~A., Lissauer, J.~J., Eke, V.~R., Sandnes, T.~D., Elphic, R.~C.\ 2025.\ Origin of Mars's moons by disruptive partial capture of an asteroid.\ Icarus 425. doi:10.1016/j.icarus.2024.116337
\bibitem[Keil et al.(1997)]{Keil1997} Keil, K., Stoeffler, D., Love, S.~G., Scott, E.~R.~D.\ 1997.\ Constraints on the role of impact heating and melting in asteroids.\ Meteoritics and Planetary Science 32, 349–363. doi:10.1111/j.1945-5100.1997.tb01278.x
\bibitem[Kleine et al.(2005)]{Kleine2005} Kleine, T., Mezger, K., Palme, H., Scherer, E., M{\"u}nker, C.\ 2005.\ Early core formation in asteroids and late accretion of chondrite parent bodies: Evidence from $^{182}$Hf- $^{182}$W in CAIs, metal-rich chondrites, and iron meteorites.\ Geochimica et Cosmochimica Acta 69, 5805–5818. doi:10.1016/j.gca.2005.07.012
\bibitem[Kobayashi et al.(2026)]{Kobayashi2026} Kobayashi, H., Tanaka, H., Hasegawa, Y., Inutsuka, S.-. ichiro .\ 2026.\ Modeling of Collisional Outcomes Based on Impact Simulations of Mars-sized Bodies.\ The Astrophysical Journal 998. doi:10.3847/1538-4357/ae1ef7
\bibitem[Kurosaki and Inutsuka(2019)]{Kurosaki2019} Kurosaki, K., Inutsuka, S.-. ichiro .\ 2019.\ The Exchange of Mass and Angular Momentum in the Impact Event of Ice Giant Planets: Implications for the Origin of Uranus.\ The Astronomical Journal 157. doi:10.3847/1538-3881/aaf165
\bibitem[Kurosaki and Inutsuka(2023)]{Kurosaki2023} Kurosaki, K., Inutsuka, S.-. ichiro .\ 2023.\ Giant Impact Events for Protoplanets: Energetics of Atmospheric Erosion by Head-on Collision.\ The Astrophysical Journal 954. doi:10.3847/1538-4357/ace9ba
\bibitem[Kurosaki and Arakawa(2026)]{Kurosaki2026} Kurosaki, K., Arakawa, M.\ 2026.\ Numerical simulation of impact cratering and induced seismic waves in sand targets.\ Icarus 446. doi:10.1016/j.icarus.2025.116874
\bibitem[Kurosawa and Genda(2018)]{Kurosawa2018} Kurosawa, K., Genda, H.\ 2018.\ Effects of Friction and Plastic Deformation in Shock-Comminuted Damaged Rocks on Impact Heating.\ Geophysical Research Letters 45, 620–626. doi:10.1002/2017GL076285
\bibitem[Leinhardt and Stewart(2012)]{Leinhardt2012} Leinhardt, Z.~M., Stewart, S.~T.\ 2012.\ Collisions between Gravity-dominated Bodies. I. Outcome Regimes and Scaling Laws.\ The Astrophysical Journal 745. doi:10.1088/0004-637X/745/1/79
\bibitem[Leinhardt and Stewart(2009)]{Leinhardt2009} Leinhardt, Z.~M., Stewart, S.~T.\ 2009.\ Full numerical simulations of catastrophic small body collisions.\ Icarus 199, 542–559. doi:10.1016/j.icarus.2008.09.013
\bibitem[Lichtenberg et al.(2019)]{Lichtenberg2019} Lichtenberg, T., Keller, T., Katz, R.~F., Golabek, G.~J., Gerya, T.~V.\ 2019.\ Magma ascent in planetesimals: Control by grain size.\ Earth and Planetary Science Letters 507, 154–165. doi:10.1016/j.epsl.2018.11.034
\bibitem[Lichtenberg et al.(2023)]{Lichtenberg2023} Lichtenberg, T., Schaefer, L.~K., Nakajima, M., Fischer, R.~A.\ 2023.\ Geophysical Evolution During Rocky Planet Formation.\ Protostars and Planets VII 534, 907. doi:10.48550/arXiv.2203.10023
\bibitem[Lucy(1977)]{Lucy1977} Lucy, L.~B.\ 1977.\ A numerical approach to the testing of the fission hypothesis.\ The Astronomical Journal 82, 1013–1024. doi:10.1086/112164
\bibitem[Mahlke et al.(2022)]{Mahlke2022} Mahlke, M., Carry, B., Mattei, P.-A.\ 2022.\ Asteroid taxonomy from cluster analysis of spectrometry and albedo.\ Astronomy and Astrophysics 665. doi:10.1051/0004-6361/202243587
\bibitem[Marcus et al.(2009)]{Marcus2009} Marcus, R.~A., Stewart, S.~T., Sasselov, D., Hernquist, L.\ 2009.\ Collisional Stripping and Disruption of Super-Earths.\ The Astrophysical Journal 700, L118–L122. doi:10.1088/0004-637X/700/2/L118
\bibitem[Melosh(1989)]{Melosh1989} Melosh, H.~J.\ 1989.\ Impact cratering : a geologic process.\ New York : Oxford University Press ; Oxford : Clarendon Press, 1989.
\bibitem[Michel et al.(2002)]{Michel2002} Michel, P., Tanga, P., Benz, W., Richardson, D.~C.\ 2002.\ Formation of Asteroid Families by Catastrophic Disruption: Simulations with Fragmentation and Gravitational Reaccumulation.\ Icarus 160, 10–23. doi:10.1006/icar.2002.6948
\bibitem[Michel et al.(2003)]{Michel2003} Michel, P., Benz, W., Richardson, D.~C.\ 2003.\ Disruption of fragmented parent bodies as the origin of asteroid families.\ Nature 421, 608–611. doi:10.1038/nature01364
\bibitem[Monaghan and Gingold(1983)]{Monaghan1983} Monaghan, J.~J., Gingold, R.~A.\ 1983.\ Shock Simulation by the Particle Method SPH.\ Journal of Computational Physics 52, 374–389. doi:10.1016/0021-9991(83)90036-0
\bibitem[Monaghan(1992)]{Monaghan1992} Monaghan, J.~J.\ 1992.\ Smoothed particle hydrodynamics..\ Annual Review of Astronomy and Astrophysics 30, 543–574. doi:10.1146/annurev.aa.30.090192.002551
\bibitem[Monaghan(1997)]{Monaghan1997} Monaghan, J.~J.\ 1997.\ SPH and Riemann Solvers.\ Journal of Computational Physics 136, 298–307. doi:10.1006/jcph.1997.5732
\bibitem[Monnereau et al.(2023)]{Monnereau2023} Monnereau, M., Guignard, J., N{\'e}ri, A., Toplis, M.~J., Quitt{\'e}, G.\ 2023.\ Differentiation time scales of small rocky bodies.\ Icarus 390. doi:10.1016/j.icarus.2022.115294
\bibitem[Moskovitz and Gaidos(2011)]{Moskovitz2011} Moskovitz, N., Gaidos, E.\ 2011.\ Differentiation of planetesimals and the thermal consequences of melt migration.\ Meteoritics and Planetary Science 46, 903–918. doi:10.1111/j.1945-5100.2011.01201.x
\bibitem[Namekata et al.(2018)]{Namekata2018} Namekata, D. and 8 colleagues 2018.\ Fortran interface layer of the framework for developing particle simulator FDPS.\ Publications of the Astronomical Society of Japan 70. doi:10.1093/pasj/psy062
\bibitem[Nesvorn{\'y} et al.(2015)]{Nesvorny2015} Nesvorn{\'y}, D., Bro{\v{z}}, M., Carruba, V.\ 2015.\ Identification and Dynamical Properties of Asteroid Families.\ Asteroids IV 297–321. doi:10.2458/azu\_uapress\_9780816532131-ch016
\bibitem[Neumann et al.(2012)]{Neumann2012} Neumann, W., Breuer, D., Spohn, T.\ 2012.\ Differentiation and core formation in accreting planetesimals.\ Astronomy and Astrophysics 543. doi:10.1051/0004-6361/201219157
\bibitem[Nichols-Fleming et al.(2024)]{Nichols-Fleming2024} Nichols-Fleming, F., Evans, A.~J., Johnson, B.~C., Sori, M.~M.\ 2024.\ Moment of Inertia and Tectonic Record of Asteroid 16 Psyche May Reveal Interior Structure and Core Solidification Processes.\ Journal of Geophysical Research (Planets) 129. doi:10.1029/2024JE008291
\bibitem[Ohnaka(1995)]{Ohnaka1995} Ohnaka, M.\ 1995.\ A shear failure strength law of rock in the brittle-plastic transition regime.\ Geophysical Research Letters 22, 25–28. doi:10.1029/94GL02791
\bibitem[Okamoto et al.(2020)]{Okamoto2020} Okamoto, T., Kurosawa, K., Genda, H., Matsui, T.\ 2020.\ Impact Ejecta Near the Impact Point Observed Using Ultra-high-Speed Imaging and SPH Simulations and a Comparison of the Two Methods.\ Journal of Geophysical Research (Planets) 125. doi:10.1029/2019JE005943
\bibitem[Palme and O'Neill(2003)]{Palme2003} Palme, H., O'Neill, H.~S.~C.\ 2003.\ Cosmochemical Estimates of Mantle Composition.\ Treatise on Geochemistry 2, 568. doi:10.1016/B0-08-043751-6/02177-0
\bibitem[Pierazzo et al. (2005)]{Pierazzo2005} Pierazzo, E., Artemieva, N. A., Ivanov, B. A. (2005). Starting conditions for hydrothermal systems underneath Martian craters: Hydrocode modeling. In T. Kenkmann, F. Hörz, \& A. Deutsch (Eds.), Large Meteorite Impacts III (Vol. 384, pp. 443–457). Geological Society of America Special Paper. https://doi.org/10.1130/0‐8137‐2384‐1.443
\bibitem[Price(2012)]{Price2012} Price, D.~J.\ 2012.\ Smoothed particle hydrodynamics and magnetohydrodynamics.\ Journal of Computational Physics 231, 759–794. doi:10.1016/j.jcp.2010.12.011
\bibitem[Raducan et al.(2020)]{Raducan2020} Raducan, S.~D., Davison, T.~M., Collins, G.~S.\ 2020.\ Morphological Diversity of Impact Craters on Asteroid (16) Psyche: Insight From Numerical Models.\ Journal of Geophysical Research (Planets) 125. doi:10.1029/2020JE006466
\bibitem[Ruiz-Bonilla et al.(2021)]{Ruiz-Bonilla2021} Ruiz-Bonilla, S., Eke, V.~R., Kegerreis, J.~A., Massey, R.~J., Teodoro, L.~F.~A.\ 2021.\ The effect of pre-impact spin on the Moon-forming collision.\ Monthly Notices of the Royal Astronomical Society 500, 2861–2870. doi:10.1093/mnras/staa3385
\bibitem[Shepard et al.(2021)]{Shepard2021} Shepard, M.~K. and 11 colleagues 2021.\ Asteroid 16 Psyche: Shape, Features, and Global Map.\ The Planetary Science Journal 2. doi:10.3847/PSJ/abfdba
\bibitem[Shuai et al.(2024)]{Shuai2024} Shuai, K., Sch{\"a}fer, C.~M., Burger, C., Hui, H.\ 2024.\ Metal-silicate mixing in planetesimal collisions.\ Astronomy and Astrophysics 687. doi:10.1051/0004-6361/202347781
\bibitem[Stewart and Leinhardt(2009)]{Stewart2009} Stewart, S.~T., Leinhardt, Z.~M.\ 2009.\ Velocity-Dependent Catastrophic Disruption Criteria for Planetesimals.\ The Astrophysical Journal 691, L133–L137. doi:10.1088/0004-637X/691/2/L133
\bibitem[Sierks et al.(2011)]{Sierks2011} Sierks, H. and 57 colleagues 2011.\ Images of Asteroid 21 Lutetia: A Remnant Planetesimal from the Early Solar System.\ Science 334, 487. doi:10.1126/science.1207325
\bibitem[Sugiura et al.(2018)]{Sugiura2018} Sugiura, K., Kobayashi, H., Inutsuka, S.\ 2018.\ Toward understanding the origin of asteroid geometries. Variety in shapes produced by equal-mass impacts.\ Astronomy and Astrophysics 620. doi:10.1051/0004-6361/201833227
\bibitem[Sugiura et al.(2020)]{Sugiura2020} Sugiura, K., Kobayashi, H., Inutsuka, S.-. ichiro .\ 2020.\ High-resolution simulations of catastrophic disruptions: Resultant shape distributions.\ Planetary and Space Science 181. doi:10.1016/j.pss.2019.104807
\bibitem[Sugiura et al.(2022)]{Sugiura2022} Sugiura, K., Haba, M.~K., Genda, H.\ 2022.\ Giant impact onto a Vesta-like asteroid and formation of mesosiderites through mixing of metallic core and surface crust.\ Icarus 379. doi:10.1016/j.icarus.2022.114949
\bibitem[Swegle et al.(1995)]{Swegle1995} Swegle, J.~W., Hicks, D.~L., Attaway, S.~W.\ 1995.\ Smoothed Particle Hydrodynamics Stability Analysis.\ Journal of Computational Physics 116, 123–134. doi:10.1006/jcph.1995.1010
\bibitem[Taylor et al.(1993)]{Taylor1993} Taylor, G.~J., Keil, K., McCoy, T., Haack, H., Scott, E.~R.~D.\ 1993.\ Asteroid Differentiation: Pyroclastic Volcanism to Magma Oceans.\ Meteoritics 28, 34. doi:10.1111/j.1945-5100.1993.tb00247.x
\bibitem[Tholen(1984)]{Tholen1984} Tholen, D.~J.\ 1984.\ Asteroid Taxonomy from Cluster Analysis of Photometry..\ Ph.D. Thesis.
\bibitem[Tillotson(1962)]{Tillotson1962} Tillotson, J.~H.\ 1962.\ Metallic Equations of State For Hypervelocity Impact.\ General Atomic Report GA-3216. 1962. Technical Report.
\bibitem[Vernazza et al.(2021)]{Vernazza2021} Vernazza, P. and 66 colleagues 2021.\ VLT/SPHERE imaging survey of the largest main-belt asteroids: Final results and synthesis.\ Astronomy and Astrophysics 654. doi:10.1051/0004-6361/202141781
\bibitem[Wei et al.(2025)]{Wei2025} Wei, Z. and 8 colleagues 2025.\ Reflectance-spectroscopic and polarization measurements of meteorite mixtures relevant to E- and M-type asteroids.\ Astronomy and Astrophysics 704. doi:10.1051/0004-6361/202555612
\bibitem[Walte et al.(2023)]{Walte2023} Walte, N.~P., Howard, C.~M., Golabek, G.~J.\ 2023.\ Mantle fragmentation and incomplete core merging of colliding planetesimals as evidenced by pallasites.\ Earth and Planetary Science Letters 617. doi:10.1016/j.epsl.2023.118247
\bibitem[Wendland (1995)]{Wendland1995} Wendland, H.\ 1995.\ Piecewise polynomial, positive definite and compactly supported radial functions of minimal degree. Adv Comput Math 4, 389–396. https://doi.org/10.1007/BF02123482
\bibitem[Zuber et al.(2022)]{Zuber2022} Zuber, M.~T. and 15 colleagues 2022.\ The Psyche Gravity Investigation.\ Space Science Reviews 218. doi:10.1007/s11214-022-00905-3

\end{thebibliography}



\section{Lists of numerical simulation results}
This appendix summarizes the whole simulation results.
\footnotesize
\begin{longtable}{cccccccccc}
    \caption{Simulation results for the impact-induced loss. \label{tab:result}}
    \\
    \hline
        Run & $M_c/M_\mathrm{T}$ &  $m_\mathrm{imp}$ & $v_\mathrm{imp}$ & $\theta$ & $Q_R$ & $M_\mathrm{ej}^M/M_\mathrm{tot}^M$ &  $M_\mathrm{ej}^C/M_\mathrm{tot}^C$ & $M_\mathrm{ej}/M_\mathrm{tot}$ & $M_b^c/M_b$ \\
    \hline
    \endfirsthead
    \caption[]{Continued from the previous page} \\
    \hline
        Run & $M_c/M_\mathrm{T}$ &  $m_\mathrm{imp}$ & $v_\mathrm{imp}$ & $\theta$ & $Q_R$ & $M_\mathrm{ej}^M/M_\mathrm{tot}^M$ &  $M_\mathrm{ej}^C/M_\mathrm{tot}^C$ & $M_\mathrm{ej}/M_\mathrm{tot}$ & $M_b^c/M_b$ \\
    \hline
    \endhead
    \hline    
         1 & 0.0 & 1.0E+23 & 2.0E+04 &  0 & 5.1E+07 & 9.3E-04 & 0.0     & 9.3E-04 & 0.0    \\
         2 & 0.0 & 1.0E+23 & 4.0E+04 &  0 & 2.0E+08 & 5.2E-02 & 0.0     & 5.2E-02 & 0.0    \\
         3 & 0.0 & 1.0E+23 & 6.0E+04 &  0 & 4.6E+08 & 2.6E-01 & 0.0     & 2.6E-01 & 0.0    \\
         4 & 0.0 & 1.0E+23 & 8.1E+04 &  0 & 8.1E+08 & 5.5E-01 & 0.0     & 5.5E-01 & 0.0    \\
         5 & 0.0 & 1.0E+23 & 1.0E+05 &  0 & 1.2E+09 & 8.2E-01 & 0.0     & 8.2E-01 & 0.0    \\
         6 & 0.0 & 1.0E+23 & 3.0E+05 &  0 & 1.1E+10 & 9.9E-01 & 0.0     & 9.9E-01 & 0.0    \\
         7 & 0.0 & 1.0E+23 & 5.0E+05 &  0 & 3.1E+10 & 1.0E+00 & 0.0     & 1.0E+00 & 0.0    \\
         8 & 0.0 & 1.0E+23 & 1.0E+06 &  0 & 1.2E+11 & 1.0E+00 & 0.0     & 1.0E+00 & 0.0    \\
         9 & 0.3 & 1.0E+23 & 2.1E+04 &  0 & 5.5E+07 & 1.3E-03 & 0.0     & 8.9E-04 & 3.0E-01\\
        10 & 0.3 & 1.0E+23 & 4.2E+04 &  0 & 2.2E+08 & 7.9E-02 & 0.0     & 5.5E-02 & 3.2E-01\\
        11 & 0.3 & 1.0E+23 & 6.3E+04 &  0 & 4.9E+08 & 4.0E-01 & 0.0     & 2.8E-01 & 4.2E-01\\
        12 & 0.3 & 1.0E+23 & 8.4E+04 &  0 & 8.7E+08 & 7.0E-01 & 1.7E-01 & 5.4E-01 & 5.5E-01\\
        13 & 0.3 & 1.0E+23 & 1.0E+05 &  0 & 1.2E+09 & 9.6E-01 & 9.0E-01 & 9.4E-01 & 5.5E-01\\
        14 & 0.3 & 1.0E+23 & 3.0E+05 &  0 & 1.1E+10 & 9.9E-01 & 9.8E-01 & 9.9E-01 & 5.8E-01\\
        15 & 0.3 & 1.0E+23 & 5.0E+05 &  0 & 3.1E+10 & 1.0E+00 & 9.9E-01 & 1.0E+00 & 7.0E-01\\
        16 & 0.3 & 1.0E+23 & 1.0E+06 &  0 & 1.2E+11 & 1.0E+00 & 1.0E+00 & 1.0E+00 & 0.0    \\
        17 & 0.5 & 1.0E+23 & 2.2E+04 &  0 & 5.8E+07 & 1.4E-03 & 0.0     & 6.9E-04 & 5.0E-01\\
        18 & 0.5 & 1.0E+23 & 4.3E+04 &  0 & 2.3E+08 & 1.1E-01 & 0.0     & 5.6E-02 & 5.3E-01\\
        19 & 0.5 & 1.0E+23 & 6.5E+04 &  0 & 5.2E+08 & 4.6E-01 & 2.5E-04 & 2.3E-01 & 6.5E-01\\
        20 & 0.5 & 1.0E+23 & 8.6E+04 &  0 & 9.3E+08 & 6.9E-01 & 2.3E-01 & 4.6E-01 & 7.1E-01\\
        21 & 0.5 & 1.0E+23 & 1.0E+05 &  0 & 1.2E+09 & 8.4E-01 & 5.3E-01 & 6.8E-01 & 7.5E-01\\
        22 & 0.5 & 1.0E+23 & 3.0E+05 &  0 & 1.1E+10 & 1.0E+00 & 9.9E-01 & 9.9E-01 & 7.8E-01\\
        23 & 0.5 & 1.0E+23 & 5.0E+05 &  0 & 3.1E+10 & 1.0E+00 & 9.9E-01 & 1.0E+00 & 8.8E-01\\
        24 & 0.5 & 1.0E+23 & 1.0E+06 &  0 & 1.2E+11 & 1.0E+00 & 1.0E+00 & 1.0E+00 & 0.0    \\
        25 & 0.7 & 1.0E+23 & 2.2E+04 &  0 & 6.2E+07 & 2.4E-03 & 0.0     & 7.2E-04 & 7.0E-01\\
        26 & 0.7 & 1.0E+23 & 4.5E+04 &  0 & 2.5E+08 & 2.2E-01 & 0.0     & 6.7E-02 & 7.5E-01\\
        27 & 0.7 & 1.0E+23 & 6.7E+04 &  0 & 5.6E+08 & 5.0E-01 & 6.9E-02 & 2.0E-01 & 8.1E-01\\
        28 & 0.7 & 1.0E+23 & 8.9E+04 &  0 & 1.0E+09 & 7.2E-01 & 3.4E-01 & 4.5E-01 & 8.5E-01\\
        29 & 0.7 & 1.0E+23 & 1.0E+05 &  0 & 1.2E+09 & 8.2E-01 & 5.1E-01 & 6.0E-01 & 8.6E-01\\
        30 & 0.7 & 1.0E+23 & 3.0E+05 &  0 & 1.1E+10 & 1.0E+00 & 9.9E-01 & 9.9E-01 & 8.5E-01\\
        31 & 0.7 & 1.0E+23 & 5.0E+05 &  0 & 3.1E+10 & 1.0E+00 & 9.9E-01 & 1.0E+00 & 1.0E+00\\
        32 & 0.7 & 1.0E+23 & 1.0E+06 &  0 & 1.2E+11 & 1.0E+00 & 1.0E+00 & 1.0E+00 & 1.0E+00\\
        33 & 1.0 & 1.0E+23 & 2.4E+04 &  0 & 7.3E+07 & 0.0     & 5.9E-04 & 5.9E-04 & 1.0E+00\\
        34 & 1.0 & 1.0E+23 & 4.8E+04 &  0 & 2.9E+08 & 0.0     & 4.4E-02 & 4.4E-02 & 1.0E+00\\
        35 & 1.0 & 1.0E+23 & 7.2E+04 &  0 & 6.5E+08 & 0.0     & 2.4E-01 & 2.4E-01 & 1.0E+00\\
        36 & 1.0 & 1.0E+23 & 9.7E+04 &  0 & 1.1E+09 & 0.0     & 5.2E-01 & 5.2E-01 & 1.0E+00\\
        37 & 1.0 & 1.0E+23 & 1.0E+05 &  0 & 1.2E+09 & 0.0     & 5.6E-01 & 5.6E-01 & 1.0E+00\\
        38 & 1.0 & 1.0E+23 & 3.0E+05 &  0 & 1.1E+10 & 0.0     & 9.9E-01 & 9.9E-01 & 1.0E+00\\
        39 & 1.0 & 1.0E+23 & 5.0E+05 &  0 & 3.1E+10 & 0.0     & 9.9E-01 & 9.9E-01 & 1.0E+00\\
        40 & 0.3 & 1.0E+21 & 1.0E+05 &  0 & 4.9E+07 & 1.0E-02 & 0.0     & 7.3E-03 & 3.0E-01\\
        41 & 0.3 & 1.0E+21 & 3.0E+05 &  0 & 4.4E+08 & 2.3E-01 & 0.0     & 1.6E-01 & 3.5E-01\\
        42 & 0.3 & 1.0E+21 & 6.0E+05 &  0 & 1.8E+09 & 6.5E-01 & 4.0E-04 & 4.6E-01 & 5.5E-01\\
        43 & 0.3 & 1.0E+21 & 1.0E+06 &  0 & 4.9E+09 & 8.5E-01 & 2.2E-01 & 6.6E-01 & 6.9E-01\\
        44 & 0.3 & 1.0E+22 & 5.0E+04 &  0 & 1.0E+08 & 2.9E-02 & 0.0     & 2.1E-02 & 2.8E-01\\
        45 & 0.3 & 1.0E+22 & 1.0E+05 &  0 & 4.1E+08 & 3.3E-01 & 0.0     & 2.4E-01 & 3.6E-01\\
        46 & 0.3 & 1.0E+22 & 4.0E+05 &  0 & 6.6E+09 & 9.8E-01 & 9.3E-01 & 9.6E-01 & 5.7E-01\\
        47 & 0.3 & 1.0E+21 & 1.0E+05 & 45 & 1.0E+07 & 1.7E-02 & 0.0     & 1.2E-02 & 3.0E-01\\
        48 & 0.3 & 1.0E+21 & 4.0E+05 & 45 & 1.6E+08 & 1.1E-01 & 0.0     & 7.9E-02 & 3.2E-01\\
        49 & 0.3 & 1.0E+21 & 7.0E+05 & 45 & 5.0E+08 & 4.1E-01 & 0.0     & 2.9E-01 & 4.2E-01\\
        50 & 0.3 & 1.0E+21 & 1.0E+06 & 45 & 1.0E+09 & 6.1E-01 & 1.7E-04 & 4.3E-01 & 5.2E-01\\
        51 & 0.3 & 1.0E+22 & 5.0E+04 & 45 & 2.1E+07 & 1.1E-01 & 0.0     & 8.3E-02 & 3.0E-01\\
        52 & 0.3 & 1.0E+22 & 1.0E+05 & 45 & 8.5E+07 & 1.6E-01 & 0.0     & 1.2E-01 & 3.1E-01\\
        53 & 0.3 & 1.0E+22 & 4.0E+05 & 45 & 1.4E+09 & 5.8E-01 & 0.0     & 4.2E-01 & 4.7E-01\\
        54 & 0.3 & 1.0E+22 & 5.0E+05 & 45 & 2.1E+09 & 6.7E-01 & 2.7E-04 & 4.9E-01 & 5.3E-01\\
        55 & 0.5 & 1.0E+21 & 5.0E+04 &  0 & 1.2E+07 & 1.1E-03 & 0.0     & 5.3E-04 & 5.0E-01\\
        56 & 0.5 & 1.0E+21 & 3.0E+05 &  0 & 4.4E+08 & 2.9E-01 & 0.0     & 1.4E-01 & 5.8E-01\\
        57 & 0.5 & 1.0E+21 & 5.0E+05 &  0 & 1.2E+09 & 5.2E-01 & 1.5E-02 & 2.7E-01 & 6.7E-01\\
        58 & 0.5 & 1.0E+21 & 6.0E+05 &  0 & 1.8E+09 & 6.3E-01 & 5.6E-02 & 3.5E-01 & 7.1E-01\\
        59 & 0.5 & 1.0E+21 & 1.0E+06 &  0 & 4.9E+09 & 8.7E-01 & 2.9E-01 & 5.9E-01 & 8.5E-01\\
        60 & 0.5 & 1.0E+22 & 3.0E+04 &  0 & 3.7E+07 & 2.2E-03 & 0.0     & 1.2E-03 & 4.5E-01\\
        61 & 0.5 & 1.0E+22 & 5.0E+04 &  0 & 1.0E+08 & 4.1E-02 & 0.0     & 2.2E-02 & 4.6E-01\\
        62 & 0.5 & 1.0E+22 & 2.0E+05 &  0 & 1.7E+09 & 5.6E-01 & 8.7E-02 & 3.4E-01 & 6.3E-01\\
        63 & 0.5 & 1.0E+22 & 5.0E+05 &  0 & 1.0E+10 & 9.7E-01 & 9.1E-01 & 9.4E-01 & 7.4E-01\\
        64 & 0.5 & 1.0E+21 & 5.0E+04 & 45 & 2.5E+06 & 9.4E-03 & 0.0     & 4.8E-03 & 5.0E-01\\
        65 & 0.5 & 1.0E+21 & 1.0E+05 & 45 & 1.0E+07 & 2.3E-02 & 0.0     & 1.2E-02 & 5.0E-01\\
        66 & 0.5 & 1.0E+21 & 3.0E+05 & 45 & 9.1E+07 & 8.7E-02 & 0.0     & 4.4E-02 & 5.2E-01\\
        67 & 0.5 & 1.0E+21 & 6.0E+05 & 45 & 3.7E+08 & 3.3E-01 & 0.0     & 1.7E-01 & 5.9E-01\\
        68 & 0.5 & 1.0E+21 & 1.0E+06 & 45 & 1.0E+09 & 5.7E-01 & 3.3E-03 & 2.9E-01 & 6.9E-01\\
        69 & 0.5 & 1.0E+22 & 3.0E+04 & 45 & 7.7E+06 & 8.9E-02 & 0.0     & 4.9E-02 & 4.8E-01\\
        70 & 0.5 & 1.0E+22 & 5.0E+04 & 45 & 2.1E+07 & 1.5E-01 & 0.0     & 8.1E-02 & 4.9E-01\\
        71 & 0.5 & 1.0E+22 & 2.0E+05 & 45 & 3.5E+08 & 3.4E-01 & 0.0     & 1.8E-01 & 5.6E-01\\
        72 & 0.5 & 1.0E+22 & 5.0E+05 & 45 & 2.1E+09 & 6.2E-01 & 2.9E-02 & 3.5E-01 & 6.8E-01\\
        73 & 0.7 & 1.0E+21 & 2.5E+04 &  0 & 3.0E+06 & 0.0     & 0.0     & 0.0     & 6.9E-01\\
        74 & 0.7 & 1.0E+21 & 1.0E+04 &  0 & 4.9E+05 & 0.0     & 0.0     & 0.0     & 6.9E-01\\
        75 & 0.7 & 1.0E+21 & 5.0E+04 &  0 & 1.2E+07 & 1.6E-03 & 0.0     & 5.0E-04 & 6.9E-01\\
        76 & 0.7 & 1.0E+21 & 3.0E+05 &  0 & 4.4E+08 & 2.4E-01 & 7.6E-04 & 7.3E-02 & 7.5E-01\\
        77 & 0.7 & 1.0E+21 & 6.0E+05 &  0 & 1.8E+09 & 6.2E-01 & 1.3E-01 & 2.8E-01 & 8.4E-01\\
        78 & 0.7 & 1.0E+21 & 1.0E+06 &  0 & 4.9E+09 & 9.9E-01 & 9.1E-01 & 9.4E-01 & 9.5E-01\\
        79 & 0.7 & 1.0E+22 & 2.3E+04 &  0 & 2.3E+07 & 7.5E-04 & 0.0     & 2.7E-04 & 6.4E-01\\
        80 & 0.7 & 1.0E+22 & 1.0E+04 &  0 & 4.1E+06 & 0.0     & 0.0     & 0.0     & 6.4E-01\\
        81 & 0.7 & 1.0E+22 & 3.0E+04 &  0 & 3.7E+07 & 3.6E-03 & 0.0     & 1.3E-03 & 6.4E-01\\
        82 & 0.7 & 1.0E+22 & 5.0E+04 &  0 & 1.0E+08 & 8.5E-02 & 0.0     & 3.1E-02 & 6.6E-01\\
        83 & 0.7 & 1.0E+22 & 2.0E+05 &  0 & 1.7E+09 & 5.6E-01 & 1.3E-01 & 2.9E-01 & 7.8E-01\\
        84 & 0.7 & 1.0E+22 & 5.0E+05 &  0 & 1.0E+10 & 9.9E-01 & 9.6E-01 & 9.7E-01 & 8.9E-01\\
        85 & 0.7 & 1.0E+21 & 2.5E+04 & 45 & 6.2E+05 & 9.7E-05 & 0.0     & 3.0E-05 & 6.9E-01\\
        86 & 0.7 & 1.0E+21 & 1.0E+04 & 45 & 1.0E+05 & 0.0     & 0.0     & 0.0     & 6.9E-01\\
        87 & 0.7 & 1.0E+21 & 1.0E+05 & 45 & 1.0E+07 & 3.8E-02 & 0.0     & 1.2E-02 & 7.0E-01\\
        88 & 0.7 & 1.0E+21 & 3.0E+05 & 45 & 9.1E+07 & 1.4E-01 & 0.0     & 4.2E-02 & 7.2E-01\\
        89 & 0.7 & 1.0E+21 & 6.0E+05 & 45 & 1.8E+08 & 3.0E-01 & 4.6E-04 & 9.4E-02 & 7.6E-01\\
        90 & 0.7 & 1.0E+21 & 1.0E+06 & 45 & 3.7E+08 & 5.9E-01 & 4.7E-02 & 2.1E-01 & 8.4E-01\\
        91 & 0.7 & 1.0E+22 & 2.3E+04 & 45 & 4.8E+06 & 8.9E-03 & 0.0     & 3.3E-03 & 6.4E-01\\
        92 & 0.7 & 1.0E+22 & 1.0E+04 & 45 & 8.5E+05 & 0.0     & 0.0     & 0.0     & 6.4E-01\\
        93 & 0.7 & 1.0E+22 & 3.0E+04 & 45 & 7.7E+06 & 1.4E-01 & 0.0     & 5.1E-02 & 6.7E-01\\
        94 & 0.7 & 1.0E+22 & 5.0E+04 & 45 & 2.1E+07 & 2.3E-01 & 0.0     & 8.2E-02 & 6.9E-01\\
        95 & 0.7 & 1.0E+22 & 2.0E+05 & 45 & 3.5E+08 & 4.1E-01 & 0.0     & 1.5E-01 & 7.5E-01\\
        96 & 0.7 & 1.0E+22 & 5.0E+05 & 45 & 2.1E+09 & 6.2E-01 & 1.1E-01 & 2.9E-01 & 8.0E-01\\
        \hline
\end{longtable}
\normalsize
    
\begin{longtable}{ccccccc}
    \caption{Analysis results for catastrophic disruption cases. We set $m_\mathrm{imp}=10^{23}$ g, $v_\mathrm{imp} = 1.0\times 10^{5}$ cm s$^{-1}$, which means that $Q_R=1.3\times 10^9$ erg g$^{-1}$. The impactor's iron core mass is equal to the target.  \label{tab:catastro}}
    \\
    \hline
        Run & $M_c/M_\mathrm{T}$ & $\theta$ & $M_{b1}/M_\mathrm{tot}$ & $M_{b1}^c/M_{b1}$ & $M_{b2}/M_\mathrm{tot}$ & $M_{b2}^c/M_{b2}$ \\
    \hline
    \endfirsthead
    \caption[]{Continued from the previous page} \\
    \hline
        Run & $M_c/M_\mathrm{T}$ & $\theta$ & $M_{b1}/M_\mathrm{tot}$ & $M_{b1}^c/M_{b1}$ & $M_{b2}/M_\mathrm{tot}$ & $M_{b2}^c/M_{b2}$ \\
    \hline
    \endhead
    \hline    
         C1 & 0.3 & 0 & 1.7E-01 & 0.59 & 5.6E-03 & 0.55  \\
         C2 & 0.3 & 10 & 1.2E-01 & 0.48 & 1.1E-01 & 0.48 \\
         C3 & 0.3 & 15 & 2.0E-01 & 0.51 & 2.0E-01 & 0.51 \\
         C4 & 0.5 &  0 & 2.2E-01 & 0.76 & 2.1E-02 & 0.73 \\
         C5 & 0.5 & 10 & 4.7E-02 & 0.76 & 4.4E-02 & 0.75 \\ 
         C6 & 0.5 & 15 & 1.6E-01 & 0.66 & 1.5E-01 & 0.65 \\
         C7 & 0.7 &  0 & 1.7E-01 & 0.87 & 1.5E-02 & 0.87 \\
         C8 & 0.7 & 10 & 3.4E-02 & 0.87 & 1.9E-02 & 0.87 \\
         C9 & 0.7 & 15 & 1.6E-02 & 0.86 & 1.1E-02 & 0.88 \\
         C10& 0.7 & 25 & 2.3E-01 & 0.80 & 2.2E-01 & 0.80 \\
         C11& 0.3 & 45 & 4.2E-01 & 0.35 & 4.2E-01 & 0.35 \\
         C12& 0.5 & 45 & 4.4E-01 & 0.57 & 4.4E-01 & 0.57 \\
         C13& 0.7 & 45 & 4.6E-01 & 0.75 & 4.6E-01 & 0.75 \\
        \hline
\end{longtable}

\begin{longtable}{ccccccc}
    \caption{The catastrophic disruption energy for the target planet with $M_\mathrm{T}=10^{23}$ g composed of an iron core $M_c$ surrounded by a rock $M_r$ mantle. This table shows the iron mass fraction ($M_c/M_\mathrm{T}$), the target radius $R_T$, the catastrophic specific impact energy for rock mantle $Q_\textrm{MD}^\ast$, and the catastrophic specific impact energy $Q_\textrm{D}^{\ast}$. In this simulation, the impactor mass is $10^{23}~$g.\label{tab:qdstar}}
    \\
    \hline
         $M_c/M_\mathrm{T}$ & $R_T$ [cm] & $Q_\textrm{MD}^\ast$ [erg g$^{-1}$] & $Q_\textrm{D}^\ast$ [erg g$^{-1}$] \\
    \hline
    \endfirsthead
    \caption[]{Continued from the previous page} \\
    \hline
         $M_c/M_\mathrm{T}$ & $R_T$ [cm] & $Q_D^M$ [erg g$^{-1}$] & $Q_D$ [erg g$^{-1}$] \\
    \hline
    \endhead
    \hline    
         0.0 & 2.1E+07 & 8.0E+08 & 8.0E+08\\
         0.3 & 1.9E+07 & 6.9E+08 & 9.5E+08\\
         0.5 & 1.8E+07 & 6.2E+08 & 1.0E+09\\
         0.7 & 1.7E+07 & 5.7E+08 & 1.1E+09\\
         1.0 & 1.5E+07  & -- & 1.1E+09 \\
        \hline
\end{longtable}
    
\begin{longtable}{cccccccc}
    \caption{Analysis results for catastrophic disruption cases considering the resolution. $N$ denotes the number of particles. \label{tab:catastro-N}}
    \\
    \hline
        Run & $M_c/M_\mathrm{T}$ & $\theta$ & $M_{b1}/M_\mathrm{tot}$ & $M_{b1}^c/M_{b1}$ & $M_{b2}/M_\mathrm{tot}$ & $M_{b2}^c/M_{b2}$ 
        &$N$  \\
    \hline
    \endfirsthead
    \caption[]{Continued from the previous page} \\
    \hline
        Run & $M_c/M_\mathrm{T}$ & $\theta$ & $M_{b1}/M_\mathrm{tot}$ & $M_{b1}^c/M_{b1}$ & $M_{b2}/M_\mathrm{tot}$ & $M_{b2}^c/M_{b2}$ & 
        $N$\\
    \hline
    \endhead
    \hline    
         C1 & 0.3 & 0 & 1.7E-01 & 0.59 & 5.6E-03 & 0.55 & $5\times 10^5$ \\
         N1 & 0.3 &  0 & 1.6E-01 & 0.59 & 1.2E-02 & 0.55 & $10^5$ \\
         N2 & 0.3 & 0 & 1.4E-01  & 0.59 & 1.8E-02 & 0.57 & $3\times 10^5$ \\  
         N3 & 0.3 & 0 & 1.3E-01 & 0.59 & 5.0E-03 & 0.56 & $10^6$ \\
        \hline
\end{longtable}

\begin{longtable}{cccccccc}
    \caption{Analysis results for catastrophic‑disruption cases are shown for various simulations: those considering the material strength of a solid mantle (Run M), both a solid mantle and solid core (Run MC), those adopting ANEOS basalt as the equation of state for the rocky mantle (Run AN), those using a pairwise formulation for artificial viscosity (Run AV), and those employing a spline kernel function (Run S). In all of these simulations, the strength of the rocky mantle is included.
\label{tab:catastro-dis}}
    \\
    \hline
        Run & $M_c/M_\mathrm{T}$ & $\theta$ & $M_{b1}/M_\mathrm{tot}$ & $M_{b1}^c/M_{b1}$ & $M_{b2}/M_\mathrm{tot}$ & $M_{b2}^c/M_{b2}$ 
        & Variation  \\
    \hline
    \endfirsthead
    \caption[]{Continued from the previous page} \\
    \hline
        Run & $M_c/M_\mathrm{T}$ & $\theta$ & $M_{b1}/M_\mathrm{tot}$ & $M_{b1}^c/M_{b1}$ & $M_{b2}/M_\mathrm{tot}$ & $M_{b2}^c/M_{b2}$ & 
        Variation\\
    \hline
    \endhead
    \hline    
         M & 0.3 & 10 & 1.1E-01  & 0.50 & 1.0E-01 & 0.89 & Solid mantle \\  
         MC & 0.3 & 10 & 6.2E-01 & 0.74 & 6.0E-04 & 0.33 & Solid core \\
         AN & 0.3 & 10 & 0.69E-01 & 0.59 & 5.0E-03 & 0.56 & ANEOS basalt \\
         AV & 0.3 & 10 & 1.6E-01 & 0.53 & 1.6E-01 & 0.89 & Pairwise AV \\
         S & 0.3 & 10 & 1.8E-01 & 0.79 & 1.3E-03 & 0.88 & Spline kernel \\
        \hline
\end{longtable}

\section*{Acknowledgement}
The authors thank the anonymous reviewers for their kind comments and valuable suggestions.
The authors thank K. Sugiura for making publicly available the SPH implementation described in \citet{Sugiura2018}, which we consulted during the development of our own simulation code.
Numerical computations were carried out on the Cray XC50 and XD2000 supercomputer at the Center for Computational Astrophysics, National Astronomical Observatory of Japan.
Our simulation code utilized FDPS.
K. Kurosaki is supported by JSPS KAKENHI Grant Numbers 25K01062, 24K07114 and 23K25927.
M. Arakawa is supported by JSPS KAKENHI Grant Number 22H00179.

\end{document}